\documentclass[runningheads]{llncs}
\usepackage{graphicx}
\usepackage{cite}

\usepackage{caption}
\usepackage[nolist]{acronym}
\usepackage{listings}
\usepackage{array}
\usepackage{bm}
\usepackage[toc,page]{appendix}
\usepackage{amsmath,amsfonts,amssymb}
\usepackage{float}
\usepackage{algpseudocode}
\usepackage[ruled,linesnumbered,noend]{algorithm2e}
\usepackage{caption}
\usepackage{subfig}
\usepackage{pgfplots}
\usepackage{tabularx}
\usepackage{multirow}
\usepackage{setspace}
\usepackage{makecell}
\usepackage{graphics}
\usepackage{svg}
\usepackage{verbatim}
\usepackage{hyphenat}
\usepackage{dsbda}
\usepackage{enumitem}

\usepackage{tikz}
\usetikzlibrary{shapes.geometric}
\usetikzlibrary{shapes.misc}
\usetikzlibrary{calc}
\usepackage{tikz-layers}

\SetCommentSty{mycommfont}

\newcommand{\summarizeAdjustedMethod}{Parallelsummarize}

\newcommand{\naiveApproach}{na\"iveApproach}

\newcommand{\BRS}{BRS\xspace} 

\newcommand{\OCtype}{OC_{\mathrm{type}}}

\newcolumntype{?}{!{\vrule width 1pt}}
\newcolumntype{L}{>{\raggedright\arraybackslash}X}
\newcolumntype{V}[1]{>{\raggedright\arraybackslash}m{#1}}
\newcolumntype{C}{>{\centering\arraybackslash}X}
\newcolumntype{M}[1]{>{\centering\arraybackslash}m{#1}}
\newcolumntype{P}[1]{>{\centering\arraybackslash}p{#1}}

\newcommand{\Cebiric}{\v{C}ebiri\'c\xspace}

\acrodefplural{DGH}[DGHs]{Domain Generalization Hierarchies}

\newlist{questions}{enumerate}{2}
\setlist[questions,1]{label=\textbf{RQ\arabic*.},ref=RQ\arabic*}
\setlist[questions,2]{label=(\alph*),ref=\thequestionsi(\alph*)}
\setlist[questions]{topsep=0pt}

\newcommand{\sig}{\mathrm{sig}}
\newcommand{\hash}{\mathrm{hash}}
\newcommand{\ID}{\mathrm{ID}}
\newcommand{\cp}{\mathrm{copy}}
    \newcommand{\studnameNQ}{Alice}
    \newcommand{\studname}{\ensuremath{\text{``\studnameNQ''}}}
    \newcommand{\studid}{\ensuremath{\mathrm{st143}}}
    \newcommand{\profid}{\ensuremath{\mathrm{pr837}}}
    \newcommand{\profnameNQ}{Bob}
    \newcommand{\profname}{\ensuremath{\text{``\profnameNQ''}}}
    \newcommand{\lectid}{\ensuremath{\mathrm{cs902}}}
    \newcommand{\lectname}{\ensuremath{\text{``Charlie''}}}
    \newcommand{\uniAid}{\ensuremath{\mathrm{xuni}}}
    \newcommand{\uniAnameNQ}{X~University}
    \newcommand{\uniAname}{\ensuremath{\text{``\uniAnameNQ''}}}
    \newcommand{\uniBid}{\ensuremath{\mathrm{uoy}}}
    \newcommand{\uniBname}{\ensuremath{\text{``Univ.~of~Y''}}}

\makeatletter
\newcommand{\manuallabel}[2]{\def\@currentlabel{#2}\label{#1}}
\makeatother

\floatstyle{plaintop}
\newfloat{ouralgorithm}{t}{loa}[section]
\floatname{ouralgorithm}{Algorithm}

\usepackage{cleveref}

\usepackage{xcolor}
\hypersetup{
    colorlinks,
    linkcolor={red!80!black},
    citecolor={blue!80!black},
    urlcolor={blue!80!black}
}

\begin{document}

\title{
Computing $k$-Bisimulations for Large Graphs: \\
A Comparison and Efficiency Analysis}

\titlerunning{Computing $k$-Bisimulations for Large Graphs}


%
\author{Jannik Rau\inst{1} 
\and
David Richerby\inst{2}\orcidID{0000-0003-1062-8451} \and \\
Ansgar Scherp\inst{1}\orcidID{0000-0002-2653-9245}}
%
%
\institute{
  University of Ulm, Germany.
  \email{\{firstname.lastname\}@uni-ulm.de}
\and
   University of Essex, UK.
  \email{david.richerby@essex.ac.uk}
}


\renewcommand{\response}[1]{#1}

\presetkeys{todonotes}{disable}{}


\maketitle 
\begin{abstract}
\shortorextended{Summarizing graphs w.r.t.\@ structural features is important to reduce the graph's size and make tasks like indexing, querying, and visualization feasible.
Our generic parallel \BRS algorithm efficiently summarizes large graphs w.r.t.\@ a custom equivalence relation~$\sim$ defined on the graph's vertices~$V$.
Moreover, the definition of~$\sim$ can be chained $k\geq 1$ times, so the defined equivalence relation becomes a $k$-bisimulation.
We evaluate the runtime and memory performance of the \BRS algorithm for $k$-bisimulation with $k=1,\ldots,10$ against two algorithms found in the literature (a sequential algorithm due to Kaushik \etal and a parallel algorithm of Schätzle \etal), which we implemented in the same software stack as \BRS.
We use five real-world and synthetic graph datasets containing 100~million to two billion edges.
\response{Our results show that the generic \BRS algorithm outperforms the respective native bisimulation algorithms on all datasets for all $k\geq5$ and for smaller~$k$ in some cases.}
The \BRS implementations of the two bisimulation algorithms run almost as fast as each other.
Thus, the \BRS algorithm is an effective parallelization of the sequential Kaushik \etal bisimulation algorithm. 
}
{Summarizing graphs w.r.t.\@ structural features is important in large graph applications, to reduce the graph's size and make tasks like indexing, querying, and visualization feasible.
The generic parallel \BRS algorithm can efficiently summarize large graphs w.r.t.\@ a custom equivalence relation~$\sim$ defined on the graph's vertices~$V$.
Moreover, $\sim$ can be chained $k\geq 1$ times, so the equivalence class of a vertex depends on the structure of the graph up to distance~$k$.
This allows the computation of stratified $k$-bisimulation, a popular concept in structural graph summarization.

In this work, we investigate whether the generic \BRS{} approach can outperform an existing single-purpose parallel algorithm for bisimulation.
Furthermore, we investigate whether an existing sequential bisimulation algorithm can effectively be computed by the parallel \BRS algorithm and how such generic implementations compete against a parallel variant.
To give a fair comparison, we have reimplemented the original algorithms in the same framework as was used for the generic \BRS algorithm.
We evaluate the performance of the two native implementations against implementations in \BRS for $k$-bisimulation with $k=1, \dots, 10$, using five real-world and synthetic graph datasets containing between $100$ million and two billion edges. 
\response{Our results show that the generic \BRS algorithm outperforms the respective native bisimulation algorithms on all datasets for all $k\geq5$ and for smaller~$k$ in some cases.}
Furthermore, the execution times of the generic \BRS algorithm for the native parallel and native sequential bisimulation variants are very similar.
This shows that the bisimulation variant computed by the native sequential algorithm can be effectively computed in parallel by the \BRS algorithm.
These insights open a new path for efficiently computing bisimulations on large graphs.}

\end{abstract} 

\section{Introduction}
\label{sec:introduction}

\extended{Large-scale graphs, for example in the form of \textit{directed, edge-labeled graphs}, are an increasingly popular way to model data~\cite{DBLP:journals/corr/abs-2003-02320}, often in the form of RDF graphs~\cite{spec/rdf-concepts-and-abstact-syntax}. These contain information about (real-word) entities, denoted by the graph's vertices, and their relationships, represented by the graph's edges.
An RDF graph can be expressed as a set of  \textit{subject}-\textit{predicate}-\textit{object} expressions.\extended{\footnote{\url{https://www.w3.org/TR/2014/REC-n-triples-20140225/Overview.html}}} These are stored as triples $(s,p,o)$, each corresponding to an edge with label~$p$ going from vertex~$s$ to vertex~$o$. 
Figure~\ref{fig:example-rdf-university} shows an example RDF graph,\footnote{For brevity, URIs are replaced by short identifiers.} which contains information about a student and the university he works at.
Such \textit{knowledge graphs}, have emerged and grown in both the open\hyp{}source and enterprise domains~\cite{DBLP:journals/corr/abs-2003-02320}.
Applications range from search engines, recommendation systems, fraud detection and business intelligence, to simple information holders.

\begin{figure}[t]
    \centering
    \subfloat[Example RDF graph for the student--university relationship]{%
        \begin{tikzpicture}[draw=black,text=black]
            \node[ellipse,draw] (stud) at (0,2) {Student};
            \node[ellipse,draw] (name) at (0,0) {\studname};
            \node[ellipse,draw] (studid) at (1.5,1) {\studid};
            \node[ellipse,draw] (uniAid) at (4,1) {\uniAid};
            \node[ellipse,draw] (uniAtype) at (6,2) {Organization};
            \node[ellipse,draw,text width=1.5cm,align=center]
                                (uniAname) at (6,0) {``University of Ulm''};

            \draw[->] (studid)-- node[midway,above right] {type} (stud);
            \draw[->] (studid)-- node[midway,below right] {name} (name);
            \draw[->] (studid)-- node[midway,above]       {worksAt} (uniAid);
            \draw[->] (uniAid)-- node[midway,above left]  {type} (uniAtype);
            \draw[->] (uniAid)-- node[midway,below left]  {name} (uniAname);
        \end{tikzpicture}
        \label{fig:example-rdf-university}
        }\\
    \subfloat[Example property graph for the student--university relationship]{%
        \begin{tikzpicture}[draw=black,text=black]
            \node[ellipse,draw] (studid) at (1.5,1) {\studid};
            \node[rounded rectangle,draw,fill=white] at (0.8,1.4) {Student};
            \node[ellipse,draw] (uniAid) at (4,1) {\uniAid};
            \node[rounded rectangle,draw,fill=white] at (5,1.4) {Organization};

            \draw[->] (studid)-- node[midway,above] (worksAt)  {worksAt} (uniAid);

            \node[rectangle,draw,text width=3cm,align=left]
                                  (worksAtProps) 
                                  at ($(worksAt) + (0,1)$)
                                  {since: ``10-09-2017''};
           \draw[dotted] (worksAt)--(worksAtProps);

            \node[rectangle,draw,align=left,anchor=north] (studentProps) at (0.5,0) {name: \studname,\\age: 23,\\...};
            \node[rectangle,draw,align=left,anchor=north] (uniProps) at (5,0) {name: \uniAname,\\founded: 1967,\\...};

            \draw[dotted] (studid)--(studentProps);
            \draw[dotted] (uniAid)--(uniProps);
        \end{tikzpicture}
        \label{fig:example-property-graph-university}
    }

    \caption{Graphs showing a student--university relationship.}
\end{figure}

An equivalent formulation is the \textit{property graph}~\cite{DBLP:journals/corr/abs-2003-02320}.
Property graphs include vertex labels, as well as property-value pairs for vertices and edges.
Figure~\ref{fig:example-property-graph-university} shows an example property graph, which contains similar information to the graph in Figure~\ref{fig:example-rdf-university}.
}
Storing, indexing, querying, and visualizing large graphs is difficult~\cite{DBLP:journals/corr/abs-2004-14794}.
One way to mitigate this challenge is \textit{graph summarization}~\cite{DBLP:journals/vldb/CebiricGKKMTZ19}.
Graphs can be summarized w.r.t.\@ so-called \emph{graph summary models}~\cite{DBLP:journals/tcs/BlumeRS21} that define structural features (e.\,g., incoming/outgoing paths), statistical measures (e.\,g., occurrences of specific vertices), or frequent patterns found in the graph~\cite{DBLP:journals/vldb/CebiricGKKMTZ19}.
This gives a \textit{summary graph}, which is usually smaller than the original but contains an approximation of or exactly the same information as the original graph w.r.t.\@ the selected features of the summary model.
Tasks that were to be performed on the original graph can be performed on the summary but much faster.
Use cases are optimizing database queries~\cite{DBLP:conf/icde/NeumannM11}, 
data visualization~\cite{DBLP:journals/vldb/GoasdoueGM20},
and OWL reasoning~\cite{DBLP:conf/dlog/VaighG21}.

\extended{This work focuses on structural graph summaries, \ie summarization w.r.t.\@ selected structural features. 
More precisely, we consider structural summaries based on equivalence relations, which form a lossless partition of the input graph's vertices.
Consequently, the summary graph contains precise structural information about selected features of the input graph~\cite{DBLP:journals/vldb/CebiricGKKMTZ19}.
Other forms of graph summarization include those based on frequent patterns or statistical measures~\cite{DBLP:journals/vldb/CebiricGKKMTZ19}, which result in approximate (lossy) summaries.
Different approaches to lossless summarization have been reported in the literature, capturing different aspects of the input graph~\cite{DBLP:journals/tcs/BlumeRS21,DBLP:journals/vldb/CebiricGKKMTZ19}.
Consequently, many \textit{single-purpose} algorithms exist, each computing one specific summary.}

Blume, Richerby, and Scherp developed a generic structural summarization approach, here referred to as \BRS~\cite{DBLP:conf/cikm/BlumeRS20,DBLP:journals/tcs/BlumeRS21}.
The \BRS algorithm summarizes an input graph w.r.t.\@ an arbitrary user-defined equivalence relation specified in its formal language FLUID~\cite{DBLP:journals/tcs/BlumeRS21}.
\extended{Researchers and practitioners in the graph domain benefit, as they do not have to rely on there being a pre-existing summary tailored to their specific needs.
Rather, they can define their custom equivalence relation and summarize graphs using the \BRS algorithm.}%
The FLUID language supports all features of structural graph summarization  found in the literature~\cite{DBLP:journals/tcs/BlumeRS21}.
There are two groups of these features.
The first comprises a vertex's \textit{local} information, \eg its label set, its direct neighbors, and the labels of its incoming or outgoing edges.
The second group considers a vertex's \textit{global} information at distance $k > 1$.
This includes, \eg local information about reachable vertices up to distance~$k$ or information about incoming or outgoing paths of length up to~$k$.
We use stratified $k$-bisimulations (formally described in Section~\ref{sec:bisim}) to summarize a graph w.r.t.\@ global information and group together vertices that have equivalent structural neighborhoods up to distance~$k$.
Several existing approaches use $k$-bisimulations to incorporate global information into structural graph summarization~\cite{DBLP:journals/tcs/BlumeRS21,DBLP:journals/vldb/CebiricGKKMTZ19}.
The \BRS algorithm generalizes these approaches and can chain any definable equivalence relation $k$~times, such that the resulting equivalence classes can be efficiently computed by global information up to distance~$k$~\cite{DBLP:journals/corr/abs-2111-12493-w-Jannik,DBLP:journals/corr/abs-2204-05821}. 

\shortorextended{}{Prior work has shown that the \BRS algorithm can effectively compute $k$-bisimulations for large values of~$k$ on real-world and synthetic graphs with up to one billion edges~\cite{DBLP:journals/corr/abs-2111-12493-w-Jannik}.}
\shortorextended{However, it~}{So far it~}is not known if a general approach like the \BRS algorithm sacrifices performance.
We choose two representative algorithms as examples to demonstrate the capabilities of our generic \BRS algorithm.
First, we have re-implemented the efficient, parallel single-purpose $k$-bisimulation algorithm of Schätzle, Neu, Lausen, and Przjaciel-Zablocki~\cite{DBLP:conf/sigmod/SchatzleNLP13} and investigate whether it outperforms our generic \BRS algorithm.
Second, we investigate the sequential algorithm for bisimulation by Kaushik, Shenoy, Bohannon, and Gudes~\cite{DBLP:conf/icde/KaushikSBG02}.
Being sequential, it is naturally disadvantaged against parallel algorithms such as ours.
However, we show in this work that the bisimulation of Kaushik \etal~\cite{DBLP:conf/icde/KaushikSBG02} can be declaratively specified and executed in the generic \BRS algorithm.
This effectively parallelizes the algorithm ``for free''.
We evaluate the performance of the \BRS-based parallelized computation of the Kaushik \etal~graph summary model and compare it with their sequential native algorithm.
We also compare both Kaushik \etal~variants, native and \BRS-based, with the parallel native algorithm of Schätzle \etal~\cite{DBLP:conf/sigmod/SchatzleNLP13} and a \BRS implementation of Schätzle \etal (see also~\cite{DBLP:journals/corr/abs-2111-12493-w-Jannik}).
Thus, we have four $k$-bisimulation algorithms.
For a fair comparison, we reimplemented the existing native algorithms in the same graph processing framework as the \BRS algorithm.
We execute the four algorithms on five graph datasets~-- two synthetic and three real-world~-- of different sizes, ranging from $100$ million edges to billions of edges.\extended{~The real-world datasets' sizes are around $100M$, $150M$, and $2B$ triples, whereas the synthetic datasets are $100M$ and $1B$ triples.} We evaluate the algorithms' performance for computing $k$-bisimulation for $k = 1, \ldots, 10$. We measure running time per iteration and the maximum memory consumption\extended{~during the computation}.
\extended{All algorithms perform computations in-memory only, to eliminate side effects.
We leave considering solutions with external memory for future work.}

\textit{The questions we address are}:
\extended{$\bullet$~}Do the native bisimulation algorithms have an advantage over a generic solution? 
\extended{$\bullet$~}How well do the native and generic algorithms scale to large real-world and synthetic graphs?
\extended{$\bullet$~}Is it possible to effectively scale a sequential algorithm by turning it into a parallel variant by using a general formal language and algorithm for graph summaries?

We discuss related work next.
Section~\ref{sec:preliminaries} defines preliminaries, while the algorithms are introduced in Section~\ref{sec:methods}.
Section~\ref{sec:experiments} outlines the experimental apparatus\extended{~including datasets, experimental procedure, implementations, and the applied measures}.
Section~\ref{sec:results} describes the results\extended{ obtained from the experiments}, and these are discussed\extended{ and interpreted} in Section~\ref{sec:discussion}. 

\section{Related Work}
\label{sec:related_work}

Summary graphs can be constructed in several ways.
\Cebiric \etal~\cite{DBLP:journals/vldb/CebiricGKKMTZ19} classify existing techniques into structural, pattern-mining, statistical, and hybrid approaches.
In this paper, we consider only structural approaches based on quotients.
Other structural summarization techniques, not based on quotients, are extensively discussed by \Cebiric \etal~\cite{DBLP:journals/vldb/CebiricGKKMTZ19}.
Structural approaches summarize a graph~$G$ w.r.t.\@ an equivalence relation~${\sim} \subseteq V \times V$ defined on the vertices~$V$ of $G$~\cite{DBLP:journals/vldb/CebiricGKKMTZ19,DBLP:journals/corr/abs-2004-14794}.
The resulting summary graph~$SG$ consists of vertices~$VS$, each of which corresponds to a equivalence class of the equivalence relation $\sim$.

\todo{Response to Reviewer \#1 with a more elaborate comparison to related work, and edge- vs vertex-based bisimulation.}
\response{One can observe that $k$-bisimulation is a popular feature for structural graph summarization~\cite{DBLP:journals/tcs/BlumeRS21}.
Bisimulation comes in three forms: backward $k$-bisimulation classifies vertices based on incoming paths of length up to~$k$, forward bisimulation considers outgoing paths, and backward-forward bisimulation considers both.
Bisimulation may be based on edge labels, vertex labels, or both, but this makes no significant difference to the algorithms.
A notion of $k$-bisimulation w.r.t.\@ graph indices is introduced by seminal works such as the $k$-RO index~\cite{DBLP:conf/icde/NestorovUWC97} and the T-index summaries~\cite{DBLP:conf/icdt/MiloS99}.
Milo and Suciu's T-index~\cite{DBLP:conf/icdt/MiloS99}, the $A(k)$-Index by Kaushik \etal~\cite{DBLP:conf/icde/KaushikSBG02}, and others are examples that summarize graphs using backward $k$-bisimulation.
\extended{Qun \etal~\cite{DBLP:conf/sigmod/QunLO03} extend the $A(k)$-Index to a $D(k)$-Index, which is also based on bisimulation but focuses on query optimization.
To this end, the $D(k)$-Index dynamically adapts its structure according to the current query load.}
We chose as representative the sequential algorithm by Kaushik \etal~\cite{DBLP:conf/icde/KaushikSBG02}, which uses vertex labels, as described in Section~\ref{sec:kaushik}.
Conversely, the $k$-RO index, the Extended Property Paths of Consens \etal~\cite{DBLP:journals/pvldb/ConsensFKP15}, the SemSets model of Ciglan \etal~\cite{DBLP:conf/www/CiglanNH12}, Buneman \etal's RDF graph alignment~\cite{DBLP:journals/pvldb/BunemanS16}, and the work of Sch\"atzle \etal~\cite{DBLP:conf/sigmod/SchatzleNLP13} are based on forward $k$-bisimulation.
We note that Sch\"atzle \etal use edge labels, as described in Section~\ref{sec:schaetzle}.
\extended{
Buneman \etal use forward $k$-bisimulation in the problem of RDF graph alignment \cite{DBLP:journals/pvldb/BunemanS16}.
Summarizing the union of two consecutive versions $G_\text{union} = G_1 \cup G_2$ of an RDF graph with respect to $k$-bisimulation, puts vertices to be aligned in the same partition.
Additionally to $k$-bisimulation, they use a similarity measure to further refine the initial $k$-bisimulation partition, as it does not capture all vertices to be aligned.
The focus of their work is the optimization of the alignment process, so that every node pair $(v_1, v_2)$, with $v_1 \in G_1$ and $v_2 \in G_2$, which have to be aligned is identified and not the construction of a $k$-bisimulation-based partition of $G$.}
\extended{Schätzle \etal compute a forward $k$-bisimulation on RDF graphs in sequential and distributed settings~\cite{DBLP:conf/sigmod/SchatzleNLP13}.
For a small synthetic dataset (${\sim}1M$ RDF-triples) the sequential algorithm slightly outperforms the distributed one; for larger datasets, the distributed algorithm clearly outperforms the sequential one.}%
Tran \etal compute a structural index for graphs based on backward-forward $k$-bisimulation~\cite{DBLP:journals/tkde/TranLR13}.
Moreover, they parameterize their notion of bisimulation to a forward-set $L_1$ and a backward-set $L_2$, so that only labels $l \in L_1$ are considered for forward-bisimulation and labels $l \in L_2$ for backward-bisimulation.
\extended{However, similar to Buneman \etal, the particular focus of their work is not the actual construction of the structural index, \eg the bisimulation partition.
Rather, they evaluate how one can efficiently optimize query processing on semi-structured data using such an index graph~\cite{DBLP:journals/tkde/TranLR13}.}
}

\extended{There are also structural summarization approaches that determine vertex equivalence only based on local information~($k \le 1$).
Campinas \etal~\cite{DBLP:conf/dexaw/CampinasPCDT12} construct summary graphs using equivalence based on outgoing edge labels and vertex labels.
With the resulting equivalence classes, known as attribute-based collections and class-based collections, they implement a query recommendation system\extended{~for SPARQL,\footnote{\url{https://www.w3.org/TR/sparql11-overview/}}} that facilitates working with heterogeneous datasets in general, and especially when the schema structure is unknown.
SchemEX~\cite{DBLP:journals/ws/KonrathGSS12} constructs a three-layered schema-level index for RDF graphs.
The third layer groups vertices $v$ and~$v'$ that have the same labels and for every edge~$(v, w)$ with label~$p$, there exists a respective $p$-labeled edge~$(v', w')$, with $w$ and $w'$ also having the same labels (and vice versa). 
The resulting equivalence classes~$[v]_{\sim}$ are mapped to the sources containing their elements.
This mapping can aid recommending related queries and generally for finding relevant data sources~\cite{DBLP:conf/kcap/GottronSKP13}.}
\extended{Several other works construct structural summaries by considering a vertex's local neighborhood \cite{DBLP:conf/webi/BenedettiBP15,DBLP:conf/www/CiglanNH12,DBLP:conf/icde/NeumannM11,DBLP:conf/esws/SpahiuPPRM16}.}

\response{Luo \etal examine structural graph summarization by forward $k$-bisimulation in a distributed, external-memory model~\cite{DBLP:conf/cikm/LuoFHWB13}.
They empirically observe that, for values of $k > 5$, the summary graph's partition blocks change little or not at all.
Therefore they state, that for summarizing a graph with respect to $k$-bisimulation, it is sufficient to summarize up to a value of $k = 5$~\cite{DBLP:conf/sigmod/LuoFHBW13}.
Finally, Martens \etal~\cite{DBLP:conf/facs2/0001GHHW21} introduce a parallel bisimulation algorithm for massively parallel devices such as GPU clusters.
Their approach is tested on a single GPU with $24\,$GB RAM, 
which limits its use on large datasets.
Nonetheless, their proposed blocking mechanism could be combined with our vertex-centric approach to further improve performance.}

\extended{
The second structural summarization technique, producing non-quotient summaries, does not use equivalence relations to summarize a graph.
Rather, the summary graph is comprised of vertex summaries~$vs$, which group together vertices~$v$ of the original graph~$G$ according to certain criteria~\cite{DBLP:journals/vldb/CebiricGKKMTZ19}. 
The main difference from quotient summaries discussed above is that, in the non-quotient summaries, a vertex~$v$ can belong to zero, one, or multiple vertex summaries~$vs$.
In contrast, in quotient summaries every vertex~$v$ has exactly one corresponding vertex summary~$VS$, which is the equivalence class of~$v$ under $\sim$~\cite{DBLP:journals/vldb/CebiricGKKMTZ19}.
Early work on non-quotient summarization includes that of Goldman and Widom ~\cite{DBLP:conf/vldb/GoldmanW97}, who created a vertex summary~$vs$ for every labeled path in the original graph~$G$.
A vertex~$v$ of $G$ is associated with a vertex summary~$vs$ if it is reachable by the corresponding label path.
The summary graph~$SG$ is used as a path-index, as well as a tool for understanding the schema structure in semi-structured databases, and hence finds application in query formulation and query optimization.
\extended{Revisiting the summarization tool SchemEX~\cite{DBLP:journals/ws/KonrathGSS12}, its first layer -- the \textit{RDF class layer} -- consists of vertex summaries~$vs_{c_j}$ representing all the classes~$c_j$ present in the input RDF graph~$G$.
A vertex~$v$ of $G$ is associated with a vertex summary~$vs_{c_j}$, iff $v$ is of the corresponding type~$c_j$.
Since a vertex~$v$ can have multiple types~$c_{j_1}, c_{j_2}, \ldots$, it is possible that~$v$ is associated with several vertex summaries~$vs_{c_{j_1}}, vs_{c_{j_2}}, \ldots$ and therefore the index's RDF class layer is considered a non-quotient summary.}
Kellou-Menouer and Kedad~\cite{DBLP:conf/er/Kellou-MenouerK15} perform schema extraction based using density-based clustering to establish a partition of the vertices based on type profiles.
For each~type $T_j$, a \textit{type profile}~$TP_j = \{(label_{1}, \alpha_{1}), (label_{2}, \alpha_{2}), \ldots \}$ is constructed, consisting of tuples of edge labels for outgoing edges~$(v, w)$ and incoming edges~$(w, v)$, with~$v \in T_j$.
The associated probabilities~$\alpha_i$ denote how likely it is that a vertex $v \in T_j$ has an edge with the respective $label_i$.
If a type profile~$TP_{j}$ contains all entries~$(label_{i}, \alpha_i)$ of another type profile~$TP_{k}$ and every~$\alpha_i$ is greater than a certain threshold~$\theta$~(\eg $\theta = 0.6$), then the vertices in~$T_{k}$ are added to~$T_{j}$ to create \textit{overlapping classes}.
Clustering can be found in more structural non-quotient approaches~\cite{DBLP:conf/aaai/UdreaPS07,DBLP:journals/computing/KhanNL15,DBLP:conf/sdm/LeFevreT10,DBLP:conf/sigmod/NavlakhaRS08}.  %

\extended{Besides structural graph summarizing, the aforementioned classification is comprised of three more categories.}
Pattern-mining approaches utilize algorithms to identify frequent patterns in the original graph~$G$, which are then used to construct the summary graph~$SG$~\cite{DBLP:journals/vldb/CebiricGKKMTZ19}.
Song \etal~\cite{DBLP:conf/icdm/SongWD16} construct \textit{$d$-summaries} to summarize a knowledge graph~$G$.
A summary~$P$, which is a graph pattern found in~$G$, is considered a $d$-summary, iff all the summary vertices~$u \in P$ are $d$-similar~($R_d$) to all their respective original vertices~$v \in V$.
Informally, $u R_d v$ iff (1)~$u$ and~$v$ share the same label and (2)~for every neighbor~$u' \in P$ of $u$ connected over an edge with label~$p$ there exists a respective neighbor~$v' \in V$ connected via the same edge label and~$u' R_{d-1} v'$.
Their definition of $d$-similarity is very similar to $k$-bisimulation (Section~\ref{sec:bisim}) and mainly differs in the domain on which it is defined, namely summary vertices and original vertices.
Statistical approaches construct summary graphs by considering quantitative properties of the input graph~$G$~\cite{DBLP:journals/vldb/CebiricGKKMTZ19}.
In example, the summarization operation $k$-SNAP~\cite{DBLP:conf/sigmod/TianHP08} minimizes a function based on occurrences of user selected edge labels to produce a summary graph~$SG$, which contains exactly $k$ vertex summaries.
In its top-down approach, it starts by partitioning the graph based on user selected vertex attributes.
Afterwards, the algorithm splits elements (vertex summaries) of the partition based on the aforementioned function, until the partition's size is~$k$.
Combining the first step, partitioning vertices by label, and the second step, minimizing a function which considers edge labels, $k$-SNAP can be considered a hybrid approach, combining structural and statistical concepts.
}

\response{Each of the works proposes a single algorithm for computing a single graph summary based on a bisimulation.
Some of the algorithms have bisimulation parameters such as the height and label parameterization in Tran \etal~\cite{DBLP:journals/tkde/TranLR13}.
We suggest a generic algorithm for computing $k$-bisimulation and show its advantages.
Also, our approach allows the bisimulation model to be specified in a declarative way and parallelizes otherwise sequential computations like in Kaushik \etal\cite{DBLP:conf/icde/KaushikSBG02} into a parallel computation.}

\section{Preliminaries}
\label{sec:preliminaries}

\extended{
We define the data structure on which the bisimulation algorithms operate in Section~\ref{sec:datastructures}.
Section~\ref{sec:bisim} defines\extended{~$k$-bisimulation as an equivalence relation on graph vertices.
Based on this, Section~\ref{sec:bisim-variants} introduces} the bisimulation models of Schätzle \etal\@~\cite{DBLP:conf/sigmod/SchatzleNLP13} and Kaushik \etal\@~\cite{DBLP:conf/icde/KaushikSBG02}.
Finally, we briefly introduce the generic graph summarization model that we use in Section~\ref{sec:gsm}, and show how the bisimulation models of Kaushik \etal and Schätzle \etal can be declaratively expressed in a formal language.}

\subsection{Data Structures}
\label{sec:datastructures}

The algorithms operate on multi-relational, labeled \shortorextended{graphs $G = (V,E, l_V, l_E)$, where $V = \{v_1, v_2, \ldots, v_n\}$ is a set of vertices and $E \subseteq V \times V$ is a set of directed edges between the vertices in~$V$.
Each vertex $v\in V$ has a finite set of labels $l_V(v)$ from a set~$\Sigma_V$ and each edge has a finite set of labels $l_E(u,v)$ from a set~$\Sigma_E$.
}%
{graphs~$G$.\footnote{We could, instead, use multi-relational property graphs, in which vertices and edges can also be labeled with key--value pairs. However, the bisimulation variants we study do not consider these key--value pairs, so multi-relational, labeled graphs suffice.}}
\shortorextended{}{\begin{definition}
\label{def:multi-relational-labeled-property-graph}
A \emph{multi-relational labeled graph}~$G$ is defined as $G = (V,E, l_V, l_E)$, where $V = \{v_1, v_2, \ldots, v_n\}$ is a set of vertices and $E \subseteq V \times V$ is a set of directed edges between the vertices in~$V$.
Furthermore, each vertex~$v \in V$ and each edge~$(u, v) \in E$ has zero or more associated labels from finite sets~$\Sigma_V$ and $\Sigma_E$, respectively.
The mapping of vertex and edge labels is done via the functions~$l_V\colon V \rightarrow \mathcal{P}(\Sigma_V)$ and $l_E\colon E \rightarrow \mathcal{P}(\Sigma_E)$, respectively.
\end{definition}}

\begin{figure*}[t]

\begin{center}

\begin{tikzpicture}[scale=0.9]
    \tikzstyle{vertex}  = [ellipse,draw]
    \tikzstyle{fvertex} = [ellipse,draw,text width=1.5cm,align=center]
    \tikzstyle{label}   = [rectangle,draw,fill=white,anchor=west]
    \tikzstyle{midlab}  = [midway,rectangle,fill=white,draw=white]

\shortorextended{
%
%
    \node[vertex] (studid) at (1.3,3.1) {\footnotesize\studid};
    \node[label] at ($(studid)+(0.1,0.45)$) {\footnotesize Student};
    \node[vertex] (profid) at (4.3,3.1) {\footnotesize\profid};
    \node[label] at ($(profid)+(0.1,0.45)$) {\footnotesize Professor};
    \node[vertex] (lectid) at (9.25,3.1) {\footnotesize\lectid};
    \node[label] at ($(lectid)+(0.1,0.45)$) {\footnotesize Lecturer};    

    \node[fvertex] (studname) at (0,1.8) {\footnotesize\studname};
    \node[label,anchor=east] at ($(studname)+(-0.3,0.45)$) {\footnotesize Literal};
    \node[vertex] (uniAid) at (2.6,1.8) {\footnotesize\uniAid};
    \node[label] at ($(uniAid)+(0.1,-0.45)$) {\footnotesize Org.};
    \node[fvertex] (profname) at (5,1.8) {\footnotesize \profname};
    \node[label] at ($(profname)+(0.3,0.45)$) {\footnotesize Literal};

    \node[vertex] (uniBid) at (8,1.8) {\footnotesize \uniBid};
    \node[label] at ($(uniBid)+(0.1,-0.4)$) {\footnotesize Org.};
    \node[fvertex] (lectname) at (10.5,1.8) {\footnotesize\lectname};
    \node[label] at ($(lectname)+(0.3,0.45)$) {\footnotesize Literal};

    \node[fvertex,text width=1.75cm](uniAname) at ($(uniAid)+(0,-1.8)$) {\footnotesize \uniAname};
    \node[label] at ($(uniAname)+(0.3,0.45)$) {\footnotesize Literal};

    \node[fvertex](uniBname) at ($(uniBid)+(0,-1.8)$) {\footnotesize \uniBname};
    \node[label] at ($(uniBname)+(0.3,0.45)$) {\footnotesize Literal};

    \newcommand{\midlabel}[1]{node[midway,rectangle,fill=white,draw=white]{\footnotesize #1}}

    \begin{scope}[on behind layer]
    \draw[-stealth] (studid)-- node[midlab] {\tiny name} (studname);
    \draw[-stealth] (studid)-- node[midlab] {\tiny worksAt}(uniAid);
    \draw[-stealth] (profid)-- node[midlab] {\tiny worksAt}(uniAid);
    \draw[-stealth] (profid)-- node[midlab] {\tiny name}(profname);
    \draw[-stealth] (lectid)-- node[midlab] {\tiny name}(lectname);
    \draw[-stealth] (lectid)-- node[midlab] {\tiny worksAt}(uniBid);
    \draw[-stealth] (uniAid)-- node[midlab] {\tiny name}(uniAname);
    \draw[-stealth] (uniBid)-- node[midlab] {\tiny name}(uniBname);
    \end{scope}
}{
%
%
    \node[vertex] (studid) at (1.3,3.4) {\studid};
    \node[label] at ($(studid)+(0.1,0.4)$) {Student};
    \node[vertex] (profid) at (4.3,3.4) {\profid};
    \node[label] at ($(profid)+(0.1,0.4)$) {Professor};
    \node[vertex] (lectid) at (11,3.4) {\lectid};
    \node[label] at ($(lectid)+(0.1,0.4)$) {Lecturer};    

    \node[fvertex] (studname) at (0,1.8) {``Jannik Rau''};
    \node[label,anchor=east] at ($(studname)+(-0.3,0.65)$) {Literal};
    \node[vertex] (uniAid) at (2.6,1.8) {\uniAid};
    \node[label] at ($(uniAid)+(-0.5,-0.45)$) {Organization};
    \node[fvertex] (profname) at (6,1.8) {``Ansgar Scherp''};
    \node[label] at ($(profname)+(0.3,0.6)$) {Literal};

    \node[vertex] (uniBid) at (9,1.8) {\uniBid};
    \node[label] at ($(uniBid)+(0.1,-0.4)$) {Organization};
    \node[fvertex] (lectname) at (13,1.8) {``David Richerby''};
    \node[label] at ($(lectname)+(0.3,0.65)$) {Literal};

    \node[fvertex](uniAname) at ($(uniAid)+(0,-2.3)$) {``University of Ulm''};
    \node[label] at ($(uniAname)+(0.3,0.65)$) {Literal};

    \node[fvertex](uniBname) at ($(uniBid)+(0,-2.3)$) {``University of Essex''};
    \node[label] at ($(uniBname)+(0.3,0.65)$) {Literal};

    \newcommand{\midlabel}[1]{node[midway,rectangle,fill=white,draw=white]{\footnotesize #1}}

    \begin{scope}[on behind layer]
    \draw[-stealth] (studid)-- node[midlab] {\footnotesize name} (studname);
    \draw[-stealth] (studid)-- node[midlab] {\footnotesize worksAt}(uniAid);
    \draw[-stealth] (profid)-- node[midlab] {\footnotesize worksAt}(uniAid);
    \draw[-stealth] (profid)-- node[midlab] {\footnotesize name}(profname);
    \draw[-stealth] (lectid)-- node[midlab] {\footnotesize name}(lectname);
    \draw[-stealth] (lectid)-- node[midlab] {\footnotesize worksAt}(uniBid);
    \draw[-stealth] (uniAid)-- node[midlab] {\footnotesize name}(uniAname);
    \draw[-stealth] (uniBid)-- node[midlab] {\footnotesize name}(uniBname);
    \end{scope}
}
\end{tikzpicture}
\end{center}

    \caption{An example graph $\boldsymbol{G}$ displaying two universities and three employees. Vertices are denoted by ellipses and edges by arrows. Vertex labels are marked with rectangles and edge labels are written on the edge.}
    \label{fig:simple_example_graph}
\end{figure*}
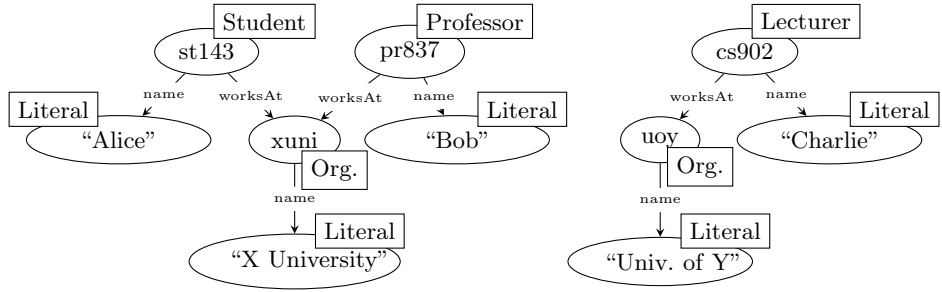

\extended{\paragraph{Example}}
Figure~\ref{fig:simple_example_graph} shows an example graph\extended{~$G = (V, E, l_V, l_E)$} 
representing two universities and three employees.
\shortorextended{%
Vertices are represented by ellipses and edges are labeled arrows. Vertex labels are shown in rectangles.}{
It has the vertices
\todo[]{produce anonymized versions}
\begin{align*}
    V & = \{ \text{asc, dri, jra, uess, uulm,} \\
      &  \text{``Ansgar Scherp'', ``David Richerby'', ``Jannik Rau'',} \\
      &  \text{``University of Essex'', ``University of Ulm''} \},
\end{align*}
and edges 
\begin{align*}
    E & = \{ \text{(asc, ``Ansgar Scherp''), (asc, uulm),} \\
      & \text{(dri, ``David Richerby''), (dri, uess),} \\
      & \text{(jra, ``Jannik Rau''), (jra, uulm),} \\
      & \text{(uess, ``University of Essex''), (uulm, ``University of Ulm'')} \}
\end{align*}
and respective vertex and edge label sets
\begin{align*}
    \Sigma_V & = \{ \text{Lecturer, Literal, Organization, Professor, Student} \}, \\
    \Sigma_E & = \{ \text{name, worksAt} \}.
\end{align*}

} 
For example, the edges~$(\profid, \uniAid)$ and $(\studid, \uniAid)$ labeled with \textit{worksAt} together with edges~$(\profid, \profname)$, $(\studid,$ $\studname)$ and $(\uniAid, \uniAname)$ labeled with \textit{name} and vertex labels~\textit{Professor}~($\profid$), \textit{Student}~($\studid$) and \textit{Organization}~($\uniAid$) state that professor \profnameNQ{} and student \studnameNQ{} both work at the organization \uniAnameNQ{}.

In a graph $G=(V,E,l_V,l_E)$ the \emph{in-neighbors} of a vertex $v\in V$ are the set of vertices $N^-(v) = \{u\mid (u,v)\in E\}$ from which $v$ receives an edge.
Similarly, $v$'s \emph{out-neighbors} are the set $N^+(v) = \{w\mid (v,w)\in E\}$ to which it sends edges.
For a set $S\subseteq V$, let $N^{+}(S) = \bigcup_{v\in S} N^{+}(v)$ be the set of out-neighbors of~$S$\extended{, \ie the vertices that have an incoming edge from some vertex in~$S$}.

\subsection{Bisimulation}
\label{sec:bisim}

A bisimulation is an equivalence relation \extended{defined }on the vertices of a directed graph~\cite{DBLP:conf/icde/KaushikSBG02,DBLP:journals/tkde/TranLR13}.
Informally, a bisimulation groups \extended{together }vertices with equivalent structural neighborhoods, \ie the neighborhoods cannot be distinguished based on the vertices' sets of labels and/or edges' labels.
\emph{Forward bisimulation (fw)} considers outgoing edges\extended{ to determine equivalence}; \emph{backward bisimulation (bw)} uses incoming edges.
Vertices $u$ and~$v$ are forward-bisimilar if, 
every out-neighbor $u'$ of~$u$
has a corresponding out-neighbor $v'$ of~$v$,
and vice versa; furthermore, the two neighbors $u'$ and~$v'$ must be bisimilar~\cite{DBLP:conf/icde/KaushikSBG02,DBLP:journals/tkde/TranLR13}.
Backward-bisimulation is defined similarly but using in-neighbors.
\shortorextended{This}{It is important to note that this} definition corresponds to a \textit{complete} bisimulation.
For the neighbors $u'$ and~$v'$ to be bisimilar, their in-/out-neighbors must be bisimilar as well.
\extended{Thus, a change to an edge in a graph could cause vertices at an arbitrary distance from that edge to become bisimilar, or to cease to be bisimilar.}%
A $k$-bi\-sim\-ul\-ation is a bisimulation on~$G$ that considers features a distance at most~$k$ from a vertex when deciding whether it is equivalent to another.

\extended{
\begin{definition}
\label{def:fw-bisim}
The \emph{forward $k$-bisimulation} ${\approx_{\mathrm{fw}}^k} \subseteq V \times V$ with $k \in \mathbb{N}$ is defined as follows:
\begin{itemize}
    \item $u \approx_{\mathrm{fw}}^0 v$ for all $u,v \in V$,
    \item $u \approx_{\mathrm{fw}}^{k+1} v$ iff $u \approx_{\mathrm{fw}}^k v$ and, for every edge $(u,u')$, there exists an edge $(v,v')$ such that $u' \approx_{\mathrm{fw}}^k v'$, and vice-versa.
\end{itemize}
\end{definition}

\begin{table}[t]
    \centering
    \small
    \begin{tabular}{|M{0.1\linewidth}|p{0.8\linewidth}|}
    \hline
         $k$ & \multicolumn{1}{c|}{Partition blocks} \\
    \hline
         0 & $V$ \\[0.5ex]
         1 & $\{ \text{asc, dri, jra, uess, uulm} \}$, $\mathrm{literals}$  \\[0.5ex]
        2 & $\{ \text{asc, dri, jra} \}$, $\{ \text{uess, uulm} \}$, $\mathrm{literals}$ \\
    \hline
    \end{tabular}
    \caption{Forward $2$-bisimulation partition of the example graph according to Definition~\ref{def:fw-bisim}.}
    \label{tab:bisim-partition-fw}
\end{table}

Table~\ref{tab:bisim-partition-fw} shows the forward bisimulation partitions of the graph in \autoref{fig:simple_example_graph}.
All vertices are $0$-bisimilar.
The $1$-bisimulation partition has two blocks: vertices that have out-neighbors, and vertices that do not.
Finally, the $2$-bisimulation partition contains three blocks: vertices with non-literal and literal out-neighbors, vertices with only literal out-neighbors, and vertices with no out-neighbors.

\begin{definition}
\label{def:bw-bisim}
The \emph{backward $k$-bisimulation} ${\approx_{\mathrm{bw}}^k} \subseteq V \times V$ with $k \in \mathbb{N}$ is defined as follows:
\begin{itemize}
    \item $u \approx_{\mathrm{bw}}^0 v$ for all $u,v \in V$,
    \item $u \approx_{\mathrm{bw}}^{k+1} v$ iff $u \approx_{\mathrm{bw}}^k v$ and, for every edge $(u',u)$ there is an edge $(v',v)$ with $u' \approx_{\mathrm{bw}}^k v'$.
\end{itemize}
\end{definition}

\begin{table}[t]
    \centering
    \small
    \begin{tabular}{|M{0.1\linewidth}|p{0.8\linewidth}|}
    \hline
         $k$ & \multicolumn{1}{c|}{Partition blocks} \\
     \hline
         0 & $V$ \\[0.5ex]
         1 & $\{ \profid, \lectid, \studid \}$, $\{\uniAid, \uniBid\} \cup \mathrm{literals}$  \\[0.5ex]
        2 &$\{\profid, \lectid, \studid\}$,
            $\{\uniAid, \uniBid, \studname,$\\
            &$\qquad\lectname, \profname\}$,\\
        &$\{ \uniBname, \uniAname \}$\\
    \hline
    \end{tabular}
    \caption{Backward $2$-bisimulation partition of the example graph according to Definition~\ref{def:bw-bisim}.}
    \label{tab:bisim-partition-bw}
\end{table}

Table~\ref{tab:bisim-partition-bw} shows the backward bisimulation partitions of the graph in Figure~\ref{fig:simple_example_graph}.
Again, all vertices are $0$-bisimilar.
This time, the $1$-bisimulation partition has two blocks: vertices with in-neighbors and vertices without.
Finally, the $2$-bisimulation partition contains three blocks: vertices with no in-neighbors, vertices whose in-neighbors have no in-neighbors, and the vertices whose in-neighbors do have in-neighbors.

\subsection{Bisimulation Variants}
}
\label{sec:bisim-variants}

The algorithms of Schätzle \etal~\cite{DBLP:conf/sigmod/SchatzleNLP13} and Kaushik \etal~\cite{DBLP:conf/icde/KaushikSBG02} compute versions of forward and backward $k$-bisimulation on labeled graphs~$G = (V,E,\Sigma_V, \Sigma_E)$\extended{, taking the labels into account}.

\extended{
\paragraph{Schätzle \etal's Forward Bisimulation}
This uses edge labels, so we modify Definition~\ref{def:fw-bisim} as follows: }

\begin{definition}
\label{def:edge-labeled-fw-bisimulation}
The \emph{edge-labeled forward $k$-bi\-sim\-ul\-ation} ${\approx_{\mathrm{fw}}^{k}} \subseteq V \times V$ with $k \in \mathbb{N}$ 
\shortorextended{of Schätzle \etal~\cite{DBLP:conf/sigmod/SchatzleNLP13}}{}
is defined as follows:
\begin{itemize}
    \item $u \approx_{\mathrm{fw}}^0 v$ for all $u,v \in V$,
    \item $u \approx_{\mathrm{fw}}^{k+1} v$ iff $u \approx_{\mathrm{fw}}^k v$ and, for every edge $(u,u')$, there is an edge $(v,v')$ with $l_E(u, u') = l_E(v, v')$ and $u' \approx_{\mathrm{fw}}^k v'$, \response{and vice-versa}.\todo{Reviewer \#2 noticed the ``vice-versa'' was missing.}
\end{itemize}
\end{definition}

\extended{
\begin{table}[t]
\scriptsize
    \centering
    \small
    \begin{tabular}{|M{0.1\linewidth}|p{0.8\linewidth}|}
    \hline
         $k$ & \multicolumn{1}{c|}{Partition blocks} \\
    \hline
         0 & $V$ \\[0.5ex]
         1 & $\{ \profid, \lectid, \studid \}$,
            $\{ \uniBid, \uniAid \}$,
            $\mathrm{literals} $ \\[0.5ex]
         2 & $\{ \profid, \lectid, \studid \}$,
            $\{ \uniBid, \uniAid \}$,
            $\mathrm{literals} $ \\
    \hline
    \end{tabular}
    \caption{Edge-labeled forward $2$-bisimulation partition of the example graph according to Definition~\ref{def:edge-labeled-fw-bisimulation}.}
    \label{tab:bisim-partition-schaetzle}

\end{table}
}

\begin{table}[t]
\small
    \centering
    \small
    \begin{tabular}{|M{0.05\linewidth}|p{0.85\linewidth}|}
    \hline
         $k$ & \multicolumn{1}{c|}{Partition blocks} \\
    \hline
         0 & $\{ \profid \}$, $\{ \lectid \}$, $\{ \studid \}$, $\{ \uniBid, \uniAid \}$, $\mathrm{literals} $  \\[0.5ex]
         1 & $\{ \profid \}$, $\{ \lectid \}$, $\{ \studid \}$, $\{ \uniBid\}$, $\{\uniAid \}$, $ \{\studname \}$, $\{\lectname \}$,\\
         &$\{\profname\}$,  
        $\{ \uniAname, \uniBname \}$  \\[0.5ex]
    2 & $\{ \profid \}$, $\{ \lectid \}$, $\{ \studid \}$, $\{ \uniBid\}$, $\{\uniAid \}$, $\{\studname \}$, $\{\lectname \}$, \\
    &$\{\profname\}$,  
        $ \{\uniAname\}$, $\{\uniBname \}$  \\
    \hline
    \end{tabular}
    
    \caption{Vertex-labeled backward $2$-bisimulation partition of the example graph according to Definition~\ref{def:vertex-labeled-bw-bisimulation}.}
    \label{tab:bisim-partition-kaushik}
\end{table}

\shortorextended{For the graph in Figure~\ref{fig:simple_example_graph},
$\approx_{\mathrm{fw}}^0$ has the single block~$V$ (as for all graphs, all vertices are initially equivalent) and 
$\approx_{\mathrm{fw}}^1$ has three blocks, \ie sets of equivalent vertices: $\{ \profid, \lectid, \studid \}$, $\{ \uniBid, \uniAid \}$, and the literals.}%
{Table~\ref{tab:bisim-partition-schaetzle} shows the edge-labeled forward $2$-bisimulation partition of the example graph in Figure~\ref{fig:simple_example_graph}.
The $2$-bisimulation is identical to the unlabeled forward $2$-bisimulation partition 
(Table~\ref{tab:bisim-partition-fw}).
However, this is reached in one step.} 
The vertices $\uniAid$ and $\uniBid$ are not $1$-bisimilar to $\studid$, $\lectid$ and $\profid$, as they have no outgoing edge labeled \textit{worksAt}. For $k\geq 2$, $k$-bisimulation in this case makes no more distinctions than $1$-bisimulation.
Note that Schätzle \etal~compute full bisimulations. 
\extended{Rather than stopping after some fixed number~$k$ of iterations, they iterate until no further changes are made to the bisimilarity relation.
For the example graph~$G$, the full bisimulation (w.r.t.\@ Definition~\ref{def:edge-labeled-fw-bisimulation}) is reached at $k = 1$.}%
We have modified their algorithm to stop after $k$~iterations.

\extended{
\paragraph{Kaushik \etal's Backward Bisimulation}
This uses vertex labels so we modify Definition~\ref{def:bw-bisim} to:}

\begin{definition}
\label{def:vertex-labeled-bw-bisimulation}
The \emph{vertex-labeled backward $k$-bi\-sim\-ul\-ation} ${\approx_{\mathrm{bw}}^{k}} \subseteq V \times V$ with $k \in \mathbb{N}$ 
\shortorextended{of Kaushik \etal~\cite{DBLP:conf/icde/KaushikSBG02}}{}
is defined as follows:
\begin{itemize}
    \item $u \approx_{\mathrm{bw}}^{0} v$ iff $l_V(u) = l_V(v)$,
    \item $u \approx_{\mathrm{bw}}^{k+1} v$ iff $v \approx_{\mathrm{bw}}^{k} u$ and, for every $(u',u)\in E$, there is $(v',v)\in E$ with $u' \approx_{\mathrm{bw}}^{k} v'$, and vice versa.
\end{itemize}
\end{definition}

Table~\ref{tab:bisim-partition-kaushik} shows the vertex-labeled backward $2$-bi\-sim\-ul\-ation partitions of the graph in Figure~\ref{fig:simple_example_graph}.
\shortorextended{$0$-bi\-sim\-ul\-ation partitions by label.}{Initially, the vertices are partitioned by label, giving the $0$-bisimulation partition.}
Then, $\uniAid$, $\uniBid$ and the \shortorextended{literals}{literal vertices} are split \shortorextended{by}{according to} their parents' labels.
No vertex is $2$-bisimilar to any other: every block is a singleton. 

\extended{Henceforth, unless otherwise specified, $\approx_{\mathrm{fw}}^{k}$ and $\approx_{\mathrm{bw}}^{k}$ will correspond to the edge-labeled forward $k$-bisimulation and the vertex-labeled backward $k$-bisimulation.}

\subsection{Graph Summaries for Bisimulation}
\label{sec:gsm}

The \BRS algorithm summarizes graphs with respect to a 
\shortorextended{\textit{graph summary model} (GSM), a mapping from graphs $G=(V,E)$ to equivalence relations~${\sim} \subseteq V \times V$.}
{
\textit{graph summary model}~\cite{DBLP:journals/tcs/BlumeRS21}.

\begin{definition}
\label{def:graph-summary-model}
A \emph{graph summary model (GSM)} is a mapping from graphs $G=(V,E)$ to equivalence relations~${\sim} \subseteq V \times V$.
\end{definition}

}
The equivalence classes of~$\sim$ partition~$G$.
A simple GSM is label equality, \ie two vertices are equivalent iff they have the same label.
Depending on the application, one might want to summarize a graph w.r.t.\@ different GSMs.
Therefore, the algorithm works with GSMs defined in our formal language FLUID~\cite{DBLP:journals/tcs/BlumeRS21}.
To flexibly and quickly define GSMs, the language provides simple and complex schema elements, along with six parameterizations, of which we use two (for details we refer to~\cite{DBLP:journals/tcs/BlumeRS21}).
\extended{The \textit{chaining parameterization} enables summarization similar to $k$-bisimulation.
The \textit{direction parameterization} determines whether incoming or outgoing edges are used.
The other four parameterizations described in~\cite{DBLP:journals/tcs/BlumeRS21} are not needed here.}

\extended{
The definitions of simple and complex schema elements, and parameterizations are based on~\cite{DBLP:journals/tcs/BlumeRS21,DBLP:journals/corr/abs-2111-12493-w-Jannik}.}

\extended{
\begin{definition}
\label{def:sse}
The three \emph{simple schema elements (SSEs)} are:
\begin{enumerate}
    \item \emph{Object cluster} (OC) compares types and vertex identifiers of all neighboring vertices: 
    two vertices $v$ and~$v'$ are equivalent iff $l_V(v) = l_V(v')$ and $N^+(v) = N^+(v')$.

\item \emph{Predicate cluster} (PC) compares labels: $v$ and~$v'$ are equivalent iff
(i) their \emph{vertex} label sets are both empty or both non-empty\footnote{This slightly unusual condition comes from the origins of FLUID in processing RDF graphs. RDF graphs do not have vertex labels; a label on vertex~$v$ is achieved by adding an edge $(v,w)$ with an edge label indicating that $w$ should be treated as a label of~$v$.} and 
(ii) they have the same labels on their outgoing edges: specifically, $\{l_E(v,w)\mid w\in N^+(v)\} = \{l_E(v',w')\mid w'\in N^+(v')\}$.

\item \emph{Predicate--object cluster} (POC) combines PC and OC: $v$ and~$v'$ are equivalent iff, $l_V(v) = l_V(v')$, $N^+(v) = N^+(v')$, and $l_E(v,w) = l_E(v',w)$ for all $w\in N^+(v)$.
\end{enumerate}
\end{definition}

The SSEs consider local information about a vertex, \ie the directly connected neighbors (``ego network'') and edge labels.
To define equivalence based on vertices at distance up to~$k$, FLUID provides complex schema elements~\cite{DBLP:journals/tcs/BlumeRS21}. 
}

\shortorextended{
A \emph{complex schema element} $CSE := (\sim_s, \sim_p, \sim_o)$ combines three equivalence relations~\cite{DBLP:journals/tcs/BlumeRS21}.
Vertices $v$ and $v'$ are equivalent, iff 
$v \sim_s v'$; and
for all $w\in N^+(v)$ there is a $w'\in N^+(v')$  with $l_E(v,w) \sim_p l_E(v',w')$ and $w \sim_o w'$, and vice versa.
}{
\begin{definition}
\label{def:cse}
A \emph{complex schema element} $CSE := (\sim_s, \sim_p, \sim_o)$ consists of three equivalence relations~\cite{DBLP:journals/tcs/BlumeRS21}.
Vertices $v$ and $v'$ are equivalent, iff 
\begin{enumerate}
\item $v \sim_s v'$,
\item for all $w\in N^+(v)$ there is a $w'\in N^+(v')$  with $l_E(v,w) \sim_p l_E(v',w')$ and 
     $w \sim_o w'$, and vice versa.
\end{enumerate}
\end{definition}
}

\extended{Introducing the identity relation~$id = \{(v,v) \mid v \in V\}$ and tautology relation~$T = V \times V$, one is able to represent the three SSEs as CSEs, \ie $OC = (T, T, id)$, $PC = (T, id, T)$, and $POC = (T, id, id)$.
An example of a CSE \extended{that takes more than local information into account~}is given by $(T, id, PC) = (T, id, (T, id, T))$.
This considers vertices as equal iff they have the same outgoing edge labels leading to neighbors with the same outgoing edge labels.}

The chaining parameterization \extended{is of special interest to us, as it }enables computing $k$-bisimulations by increasing the neighborhood considered for determining vertex equivalence~\cite{DBLP:journals/tcs/BlumeRS21}.
It is defined by nesting CSEs.
\shortorextended{Given a complex schema element $CSE := ({\sim_s,} {\sim_p,} \sim_o)$ and $k\in\mathbb{N}_{>0}$, the chaining parameterization $CSE^k$ defines the equivalence relation that corresponds to recursively applying CSE to a distance of $k$~hops. $CSE^{1} := (\sim_s, \sim_p, \sim_o)$ and, inductively for $k>1$, $CSE^{k} := (\sim_s, \sim_p,$ $CSE^{k-1})$.}{
\begin{definition}
\label{def:chaining-parameterization}
The \emph{chaining parameterization} $cp(CSE, k)$ takes as input a complex schema element $CSE := (\sim_s, \sim_p, \sim_o)$ and a chaining parameter~$k \in \mathbb{N}_{>0}$ and returns an equivalence relation~$CSE^k$ that corresponds to recursively applying CSE to a distance of $k$~hops.
$CSE^k$ is defined inductively as
\begin{align*}
    CSE^{1} & = (\sim_s, \sim_p, \sim_o), \\
    CSE^{k} &= (\sim_s, \sim_p, CSE^{k-1}) \text{ for } k > 1\,.
\end{align*}
\end{definition}
}
\todo{Response to Reviewer \#1 explaining the quotient graph (=summary).}
\response{This results in a summary graph that has one vertex for each equivalence class in each equivalence relation defined within the CSE.
Summary vertices $v$ and~$w$ are connected via a labeled edge, if all vertices in the input graph represented by~$v$ have an edge with this label to a vertex in~$w$.
%
For full details, see~\cite{DBLP:journals/tcs/BlumeRS21}.}

To model backward $k$-bisimulations, we need to work with incoming edges, whereas the\extended{~simple and complex} schema elements consider only outgoing edges. In FLUID, this is done with the direction parameterization~\cite{DBLP:journals/tcs/BlumeRS21} but, here, we simplify notation. 
\shortorextended{We write $SE^{-1}$ for the schema element defined analogously to~$SE$ but using the relation $E^{-1} = \{(y,x)\mid (x,y)\in E\}$ in place of the graph's edge relation~$E$, and the edge labeling $\ell^{-1}_E(y,x) = \ell_E(x,y)$.}{
\begin{definition}
\label{def:inverse-schema-element}
For any \extended{simple or~}complex schema element~$SE$, the \emph{inverse schema element} $SE^{-1}$ is defined analogously to $SE$ but using the relation $E^{-1} = \{(y,x)\mid (x,y)\in E\}$ in place of the graph's edge relation~$E$, and the edge labeling $\ell^{-1}_E(y,x) = \ell_E(x,y)$..
In particular, $(\sim_s,\sim_p,\sim_o)^{-1}$ denotes the inverse of the CSE $(\sim_s,\sim_p,\sim_o)$.
\end{definition}
} %

\extended{Using the chaining parameterization, the edge-labeled forward $k$-bisimulation of Schätzle \etal (see \cite{DBLP:conf/sigmod/SchatzleNLP13} and Definition~\ref{def:edge-labeled-fw-bisimulation}) can be defined.
Using chaining and the inverse schema element definition allows us to define the vertex-labeled backward $k$-bisimulation of Kaushik \etal (see \cite{DBLP:conf/icde/KaushikSBG02} and Definition~\ref{def:vertex-labeled-bw-bisimulation}).}
\response{Following Definitions \ref{def:edge-labeled-fw-bisimulation} and~\ref{def:vertex-labeled-bw-bisimulation} of Schätzle \etal and Kaushik \etal, we define $CSE_{\text{Sch}}$ and $CSE_{\text{Kau}}$ as follows.}
Here, $id = \{(v,v) \mid v \in V\}$ and $T = V \times V$, and vertices are equivalent in $\OCtype$ iff they have the same labels~\cite{DBLP:journals/tcs/BlumeRS21}.
\begin{align}
\label{gsm:schaetzle}
        CSE_{\text{Sch}} &:= (T, id, T)^k \\
\label{gsm:kaushik} 
        CSE_{\text{Kau}} &:= \big( (\OCtype, T, \OCtype)^{-1}\big)^k \,. 
\end{align}

\extended{Vertices are equivalent in $\OCtype$ iff they have the same labels~\cite{DBLP:journals/tcs/BlumeRS21}.}

\section{Algorithms}
\label{sec:methods}

We introduce the single-purpose algorithms of Schätzle \etal~\cite{DBLP:conf/sigmod/SchatzleNLP13} in Section~\ref{sec:schaetzle} and Kaushik \etal~\cite{DBLP:conf/icde/KaushikSBG02} in Section~\ref{sec:kaushik}.
Finally, we introduce the generic \BRS algorithm in Section~\ref{sec:brs}. 
The first two algorithms compute summaries in a fundamentally different way to the \BRS algorithm.
At the beginning of the execution, \BRS considers every vertex to be in its own equivalence class.
During execution, vertices with the same vertex summary are merged.
Therefore, \BRS can be seen as a \textit{bottom-up} approach.
In contrast, Schätzle \etal~consider all vertices to be equivalent at the beginning, and Kaushik \etal~initially consider all vertices with the same label to be equivalent.
The equivalence relation is then successively refined.
These two algorithms can be seen as \textit{top-down} approaches.

Here, it is convenient to consider set partitions.
A partition of a set~$V$ is a set $\{B_1, \dots, B_\ell\}$ such that: (i) $\emptyset\subsetneq B_i\subseteq V$ for each~$i$, (ii) $\bigcup_i B_i = V$, and (iii) $B_i\cap B_j = \emptyset$ for each $i\neq j$.
The sets $B_i$ are known as \emph{blocks}.
The equivalence classes of an equivalence relation over~$V$ partition the graph's vertices.
A key concept is \emph{partition refinement}.
\shortorextended{A partition $P_i = \{B_{i1}, B_{i2}, \ldots \}$ \emph{refines} $P_j = \{B_{j1}, B_{j2}, \ldots \}$ iff every block $B_{ik}$ of~$P_i$ is contained in a block $B_{j\ell}$ of~$P_j$.
}{
\begin{definition}
\label{def:partition-refinement}
Let $P_i = \{B_{i1}, B_{i2}, \ldots \}$ and $P_j = \{B_{j1}, B_{j2}, \ldots \}$ be two partitions of a finite set~$V$. 
$P_i$~is a \emph{refinement} of~$P_j$ if every block $B_{ik}$ of~$P_i$ is contained in a block $B_{j\ell}$ of~$P_j$. 
\end{definition}}

\extended{In Table~\ref{tab:comparison_of_algorithms}, the considered algorithms are compared concerning their \textit{paradigm} (\eg parallel vs. sequential), \textit{origin} (which work they are based on), \textit{worst-case complexity}, and the computed \textit{graph summary model}.
The following subsections describe the algorithms' origins and procedure in more detail.

\begin{table*}[t]
    \centering
    \small
    \begin{tabular}{|M{0.15\linewidth}M{0.15\linewidth}M{0.2\linewidth}M{0.2\linewidth}M{0.15\linewidth}|}
    \hline
         \textbf{Algorithm} & \textbf{Paradigm} & \textbf{Based on} &  \textbf{Worst-Case Complexity} & \textbf{Graph Summary Model} \\
         \hline
         \multicolumn{5}{l}{Set Union Approaches (Bottom-Up)} \\
         \hline
         \BRS \cite{DBLP:conf/cikm/BlumeRS20} & Parallel -- Signal/Collect & Set Union Problem \cite{DBLP:journals/jacm/TarjanL84} & \( \mathcal{O}(n + m \cdot \alpha(m+n,n))\) & \textit{Generic}, Section~\ref{sec:gsm} \\
         \hline
         \hline
        \multicolumn{5}{l}{Set Refinement Approaches (Top-Down)} \\
                  
         \hline
         Schätzle \etal~\cite{DBLP:conf/sigmod/SchatzleNLP13} & Distributed -- MapReduce & Blom and Orzan \cite{DBLP:journals/entcs/BlomO02}, a distributed version of Kanellakis and Smolka \cite{DBLP:journals/iandc/KanellakisS90} & \( \mathcal{O}(m \cdot n + n^{2})\) & Specific, Definition~\ref{def:edge-labeled-fw-bisimulation}, \eqref{gsm:schaetzle} \\
         \hline
         Kaushik \etal~\cite{DBLP:conf/icde/KaushikSBG02} & Sequential & Paige and Tarjan \cite{DBLP:journals/siamcomp/PaigeT87} & \( \mathcal{O}(k \cdot m)\) & Specific, Definition~\ref{def:vertex-labeled-bw-bisimulation}, \eqref{gsm:kaushik} \\
     \hline
    \end{tabular}
    \caption{Comparison of the considered algorithms.}
    \label{tab:comparison_of_algorithms}
\end{table*}
}

\subsection{Native Schätzle \etal~Algorithm}
\label{sec:schaetzle}
\shortorextended{This algorithm~\cite{DBLP:conf/sigmod/SchatzleNLP13} is a distributed MapReduce approach for reducing labeled transition systems.}
{The approach taken by Schätzle \etal~\cite{DBLP:conf/sigmod/SchatzleNLP13} is distributed and uses the \textit{MapReduce} paradigm.
It is based on an implementation of Blom and Orzan~\cite{DBLP:journals/entcs/BlomO02} (see Table~\ref{tab:comparison_of_algorithms}) for reducing labeled transition systems~(LTS), a form of a labeled directed graph, modulo strong bisimulation.
Blom and Orzan's implementation, in turn, is a distributed version of the ``na\"ive'' method established by Kanellakis and Smolka~\cite{DBLP:journals/iandc/KanellakisS90}.
Kanellakis and Smolka examine the problem of testing observational equivalence on Calculus of Communicating Systems~(CCS) expressions, which can be represented as labeled directed graphs~\cite{DBLP:journals/iandc/KanellakisS90}.
Their definition of $k$-limited observational equivalence corresponds to a stratified $k$-bisimulation.

Two differences between the work of Schätzle \etal and Blom and Orzan is that Schätzle \etal use RDF graphs and the Apache Hadoop Framework.\footnote{https://hadoop.apache.org/} This is a more modern framework than that of Blom and Orzan, who use the Message Passing Interface\footnote{\url{https://www.mpi-forum.org/index.html}} to compute $k$-bisimulations of labeled transition systems.

The approach of Schätzle \etal is detailed in this section, though the concepts are nearly the same for Blom and Orzan's work.
We have adapted the definitions and pseudocode to multi-relational labeled property graphs~(Definition~\ref{def:multi-relational-labeled-property-graph}).

}
Two fundamental concepts are the \textit{signature} and  \textit{ID} of a vertex~$v$ with respect to the current iteration's partition~$P_i$.

\shortorextended{
The \emph{signature} of a vertex $v$ w.r.t.\@ a partition $P_i = \{B_{i1}, B_{i2}, \ldots \}$ of~$V$ is given by
$\sig_{P_i}(v) = \{(\ell, B_{ij}) \mid (v, w) \in E \text{ with } l_E(v, w) = \ell \text{ and } w \in B_{ij} \}$.
}{
\begin{definition}[\cite{DBLP:conf/sigmod/SchatzleNLP13}]
\label{def:signature}
The \emph{signature} of a vertex $v$ with respect to a partition $P_i = \{B_{i1}, B_{i2}, \ldots \}$ of~$V$ is given by
\begin{align*}
    \sig_{P_i}(v) = \{(\ell, B_{ij}) \mid (v, w) \in E \text{ with } l_E(v, w) = \ell \text{ and } w \in B_{ij} \}\,.
\end{align*}
\end{definition}

}That is, $v$'s signature w.r.t.\@ the current iteration's partition~$P_i$ is the set of outgoing edge labels to blocks of $P_i$.
By Definition~\ref{def:edge-labeled-fw-bisimulation}, \extended{vertices $u$ and~$v$ are $(k+1)$-bisimilar, }$u \approx_{\mathrm{fw}}^{k+1} v$, iff $\sig_{P_k}(u) = \sig_{P_k}(v)$.
Therefore signatures identify the block of a vertex and represent the current bisimulation partition, and the signature of~$v$ w.r.t.~$P_i$ can be represented as
\shortorextended{%
$\sig_{P_{i+1}}(v) = \{(l_E(v, w), \sig_{P_i}(w)) \mid (v, w) \in E\}$.}%
{
\[
    \sig_{P_{i+1}}(v) = \{(\ell, \sig_{P_i}(w)) \mid (v, w) \in E \text{ with } l_E(v, w) = \ell\}\,.
\]
We now introduce the $\ID$ function.}
\todo{Response to Reviewer \#1 comment on citing Hellings et al., SIGMOD 2012, to introduce the idea of working with hash values.

As well how the $\ID$ function is defined.
}
\response{The nested structure of vertex signatures means they can become very large.
Thus, we compute a recursively defined hash value proposed by Hellings \etal~\cite{DBLP:conf/sigmod/HellingsFH12}, which is also used by Schätzle \etal~\cite{DBLP:conf/sigmod/SchatzleNLP13}.
We use this hash function to assign $\sig_{P_i}(v)$ an integer value which we denote by $\ID_{P_i}(v)$\extended{~(see also discussion on hash-based messaging in~\cite{DBLP:journals/corr/abs-2111-12493-w-Jannik})}.}
\extended{
\begin{definition}
\label{def:id}
The \emph{ID} of a vertex~$v$ with respect to a partition~$P_i$ is the function $\ID_{P_i}\colon V \rightarrow \mathbb{N}$ defined by
$\ID_{P_i}(v) = \hash(\sig_{P_i}(v))$,
which computes a hash value for $\sig_{P_i}(v)$.
\end{definition}}%
Now the signature of a vertex~$v$ w.r.t.\@ the current partition~$P_i$ can be represented as
\shortorextended{$\sig_{P_{i+1}}(v) = \{(l_E(v, w), \ID_{P_i}(w)) \mid (v, w) \in E\}$.
}{
\[
    \sig_{P_{i+1}}(v) = \{(\ell, \ID_{P_i}(w)) \mid (v, w) \in E \text{ with } l_E(v, w) = \ell\}\,.
\]
}

With $\sig_{P_i}(v)$ and $\ID_{P_i}(v)$, the procedure for computing an edge-labeled forward $k$-bisimulation partition is outlined in~Algorithm~\ref{alg:schaetzle-alg}.
The initial partition is just~$V$ (line~\ref{algo:initialisation-schaetzle}), as every vertex is $0$-bisimilar to every other vertex.
Next, the algorithm performs $k$~iterations (lines \ref{algo:schaetzle-loop-start}--\ref{algo:schaetzle-loop-end}). 
In the $i$th iteration, the information needed to construct a vertex's signature $\sig_{P_i}(v)$ is sent to every vertex~$v$ (line~\ref{algo:map-job}). 
This information is the edge label $l_E(v, w)\in \Sigma_E$ and the block identifier~$\ID_{P_{i-1}}(w)$ for every $w\in N^+(v)$. 
The signature~$\sig_{P_i}(v)$ is then constructed using the received information, and the identifiers~$\ID_{P_i}(v)$ are updated for all~$v$~(line~\ref{algo:reduce-job}).
At the end of each iteration, the algorithm checks if any vertex ID was updated, by comparing the number of distinct values in $\ID_{P_i}$ and $\ID_{P_{i-1}}$~(line~\ref{algo:check-count}).\extended{\footnote{Since blocks can be split but not merged, it suffices to count the blocks.}}
If no vertex ID was updated, we have reached full bisimulation~\cite{DBLP:journals/entcs/BlomO02,DBLP:conf/sigmod/SchatzleNLP13} and hence can stop execution early. 
At the end, the resulting $k$-bisimulation partition~$P_k$ is constructed by putting vertices~$v$ in one block if they share the same identifier value~$\ID_{P_i}(v)$ (line~\ref{algo:merge-vertices-schaetzle}).

%
%
\begin{algorithm}
\small
\SetAlgoLined
\SetFuncSty{textsc}
\SetKwProg{Fn}{function}{}{end}
\SetKwInput{Input}{Input}
\SetKw{Result}{returns}
\SetKw{KwBreak}{break}
\SetKwFunction{bisim}{bisimSchätzle}
\SetNoFillComment
\caption{Bisimulation Algorithm by Schätzle \etal~\cite{DBLP:conf/sigmod/SchatzleNLP13}}
\label{alg:schaetzle-alg}
\Fn{\bisim{$G = (V,E, l_V, l_E)$, $k \in \mathbb{N}$}}{
    \tcc{Initially, all $v\in V$ in same block with $\ID_{P_0}(v) = 0$}
    $P_0 \gets \{V\}$\;\label{algo:initialisation-schaetzle}
    \For{$i\gets 1$ \KwTo $k$}%
    {\label{algo:schaetzle-loop-start}
        \tcc{Map Job}
        \For{$(v,w) \in E$}
          {Send $(l_E(v, w), \ID_{P_{i-1}}(w))$}\label{algo:map-job}
        \vspace{1ex}
        \tcc{Reduce Job}
        Construct $\sig_{P_i}(v)$ and update $\ID_{P_i}(v)$\;\label{algo:reduce-job}
        \vspace{1ex}
        \tcc{Check if full bisimulation is reached}
        \If{\label{algo:check-count}$|\ID_{P_i}| = |\ID_{P_{i-1}}|$}%
        {
            \KwBreak\;
        }
    }\label{algo:schaetzle-loop-end}
    \vspace{1ex}
    Construct $P_k$ from $\ID_{P_i}$\;\label{algo:merge-vertices-schaetzle}
    \Return $P_k$\;
}
\end{algorithm}


\subsection{Native Kaushik \etal Algorithm}
\label{sec:kaushik}

\shortorextended{This algorithm~\cite{DBLP:conf/icde/KaushikSBG02}}{In contrast to the Schätzle \etal algorithm, the native Kaushik \etal~algorithm\cite{DBLP:conf/icde/KaushikSBG02}} sequentially computes vertex-labeled backward $k$-bisimulations.
\extended{It is an adapted version of Paige and Tarjan's ``na\"ive method''~\cite{DBLP:journals/siamcomp/PaigeT87} for solving the \textit{relational coarsest partition problem}.
The notation used by Kaushik \etal and Paige and Tarjan are clarified such that they are uniform for both, which is not the case in the original papers.
The details can be found in Table~\ref{tab:notions} in Appendix \ref{sec:appendix-mappings}. }%
The following definitions are \shortorextended{from}{due to Paige and Tarjan}~\cite{DBLP:journals/siamcomp/PaigeT87}, 
\shortorextended{modified for backward bisimulation}
{but we have modified them to deal with backward instead of forward bisimulation}.

\shortorextended{A subset $B \subseteq V$ is \emph{stable} with respect to another subset $S \subseteq V$ if either  $B \subseteq N^{+}(S)$ or $B \cap N^{+}(S) = \emptyset$.}{
\begin{definition}[Stable subset]
\label{def:subset-stable}
A subset $B \subseteq V$ is \emph{stable} with respect to another subset $S \subseteq V$ if either  $B \subseteq N^{+}(S)$ or $B \cap N^{+}(S) = \emptyset$.
\end{definition}
}
That is, vertices in a stable set~$B$ are indistinguishable by their relation to~$S$: either all vertices in~$B$ get at least one edge from~$S$, or none do.
\shortorextended{If $B$~is not stable w.r.t.~$S$, we call $S$ a \emph{splitter} of~$B$.}{}

\shortorextended{Building on this, partition $P_i$ of~$V$ is \emph{stable with respect to a subset $S \subseteq V$} if every block $B_{ij} \in P_i$ is stable w.r.t.~$S$.
$P_i$~is \emph{stable} if it is stable w.r.t.\@ each of its blocks $B_{ij}$.}{
\begin{definition}[Stable partition]
\label{def:partition-subset-stable}
\label{def:partition-stable}
A partition $P_i$ of~$V$ is \emph{stable} with respect to a subset $S \subseteq V$ if every block $B_{ij} \in P_i$ is stable w.r.t.~$S$.
$P_i$~is stable if it is stable w.r.t.\@ each of its blocks $B_{ij}$.
\end{definition}
}
Thus, a partition $P_i$ is stable if none of its blocks $B_{ij}$ can be split into a set of vertices that receive edges from some~$B_{ik}$ and a set of vertices that do not. A stable partition corresponds to the endpoint of a bisimulation computation: no further distinctions can be made.

\shortorextended{This gives an algorithm for bisimulation, due to Paige and Tarjan~\cite{DBLP:journals/siamcomp/PaigeT87} who refer to it as the ``na\"ive algorithm''. The initial partition is repeatedly refined by using its own blocks or unions of them as splitters: if $S$~splits a block~$B$, we replace $B$ in the partition with the two new blocks $B\cap N^+(S)$ and $B-N^+(S)$.}%
{\begin{definition}
\label{def:rel-coar-par-prob}
In the \emph{relational coarsest partition problem}, we are given a graph $G=(V,E,l_V,l_E)$ and an initial partition $P_0$ of~$V$. We must find the \emph{coarsest stable refinement} of $P_0$, the unique stable partition~$P$ such that $P$ is a refinement of~$P_0$ and every other stable partition is a refinement of~$P$.
\end{definition}

Among all stable refinements of~$P_0$, the coarsest stable refinement has the fewest blocks. The na\"ive method of Paige and Tarjan for solving the relational coarsest partition problem is defined in Algorithm~\ref{alg:ptnaive-alg}. 
It uses of the $\mathrm{split()}$ function. This function's arguments are a partition~$P$ and a vertex set~$S$, which we call a ``splitter''. $\mathrm{split}(P,S)$ replaces every block~$B$ of~$P$ that is not stable w.r.t.~$S$ with the two sub-blocks consisting, respectively, of vertices in~$B$ that send edges to~$S$, and those that do not.

\extended{
\begin{definition}[Split function]
\label{def:split-original}
For any partition $P_i$ and set $S \subseteq V$, let $\mathrm{split}(S,P_i)$ be the refinement of $P_i$ obtained by replacing each block $B_{ij} \in P_i$ that is not stable w.r.t.~$S$, by the blocks $B_{ij} \cap N^{+}(S)$ and $B_{ij} - N^{+}(S)$. We refer to the set~$S$ as a \emph{splitter}.
\end{definition}}

This leads to Algorithm~\ref{alg:ptnaive-alg}, which repeatedly refines the initial partition
by using its own blocks or unions of them as splitters. }
When no more splitters exist, the partition is stable~\cite{DBLP:journals/siamcomp/PaigeT87}
and equivalent to the full backward bisimulation of the initial partition~$P_0$~\cite{DBLP:journals/iandc/KanellakisS90}.
\extended{\footnote{Kanellakis and Smolka refer to this as \emph{observational equivalence} or \emph{strong equivalence.}}}

\shortorextended{}{\begin{algorithm}
\scriptsize
\SetAlgoLined
\SetFuncSty{textsc}
\SetKwProg{Fn}{function}{}{end}
\SetKwInput{Input}{Input}
\SetKw{Result}{returns}
\SetKwFunction{naive}{\naiveApproach{}}
\SetNoFillComment
\caption{Na\"ive algorithm for relational coarsest partition of initial partition $P_0$ of $V$}
\label{alg:ptnaive-alg}
\Fn{\naive{$P_0$ }}{
    $i \gets 0$\;
\While{\textup{$\exists$ a splitter $S$, which is a union of blocks $B_{ij} \in P_i$}}{
    $i \gets i+1$\;
    $P_i \gets \mathrm{split}(S,P_{i-1})$\;
}
\Return $P_i$\;
}
\end{algorithm}
}

The algorithm of Kaushik \etal, Algorithm~\ref{alg:kaushik-alg}, modifies \shortorextended{this na\"ive approach}{Algorithm~\ref{alg:ptnaive-alg}}.
The first difference is that in each iteration~$i \in \{1, \ldots, k\}$ the partition is stabilized with respect to each of its own blocks~(lines \ref{algo:mid-loop-kaushik}--\ref{algo:mid-loop-kaushik-end}).
This ensures that, after iteration~$i$, the algorithm has computed the $i$-bisimulation \cite{DBLP:conf/icde/KaushikSBG02}, which is not the case in \shortorextended{the na\"ive algorithm}{Algorithm~\ref{alg:ptnaive-alg}}.
Second, blocks are split as \shortorextended{defined above}{in Definition~\ref{def:split-original}}, (lines \ref{algo:kaushik-set-succ}--\ref{algo:split-end}).
As a result, Algorithm~\ref{alg:kaushik-alg} computes the $k$-backward bisimulation.
To check if \shortorextended{full bisimulation has been reached}{$P$ is the relational coarsest partition of $P_0$ (i.e., if full bisimulation has been reached)}, the algorithm uses the Boolean variable $\mathrm{wasSplit}$~(lines~\ref{algo:wasSplit-initialise} and~\ref{algo:wasSplit-update}).
If this is~$\mathrm{false}$ at the end of an iteration~$i$, no block was split, so the algorithm stops early~(line~\ref{algo:wasSplit-check}).
Moreover, Algorithm~\ref{alg:kaushik-alg} tracks which sets have been used as splitters (line~\ref{algo:usedSplitters}), to avoid checking for stability against sets w.r.t.\@ which the partition is already known to be stable.
The algorithm provided by Kaushik \etal does not include this.
If a partition~$P$ is stable w.r.t.\@ a block~$B$, each refinement of~$P$ is also stable w.r.t.\@ $B$~\cite{DBLP:journals/siamcomp/PaigeT87}.
So after the partition~$P$ is stabilized w.r.t.\@ a block copy~$B^{\cp}$~(lines \ref{algo:mid-loop-kaushik}--\ref{algo:mid-loop-kaushik-end}), we can add $B^{\cp}$ to the $\mathrm{usedSplitters}$ set and not consider it in subsequent iterations.


\begin{algorithm}[!t]
\small
\SetAlgoLined
\SetFuncSty{textsc}
\SetKwProg{Fn}{function}{}{end}
\SetKwInput{Input}{Input}
\SetKw{Result}{returns}
\SetKwFunction{bisim}{bisimKaushik}
\SetKw{KwAnd}{and}
\SetKw{KwBreak}{break}
\SetNoFillComment
\caption{Bisimulation Algorithm by Kaushik \etal~\cite{DBLP:conf/icde/KaushikSBG02}}
\label{alg:kaushik-alg}
\Fn{\bisim{$G = (V,E, l_V, l_E)$, $k \in \mathbb{N}$}}{
    \tcc{$P := \{ B_{1}, B_{2}, \ldots, B_{t} \}$}
    $P \gets $ partition $V$ by label\;  
    $\mathrm{usedSplitters} \gets \emptyset$\;\label{algo:usedSplitters}
\For{$i\gets 1, \dots, k$}{ \label{algo:stabilizie-routine-start}
    $P^{\cp} \gets P$\;
    $\mathrm{wasSplit} \gets \mathrm{false}$\;\label{algo:wasSplit-initialise}
    \For{$B^{\cp} \in P^{\cp} - \mathrm{usedSplitters}$}{ \label{algo:mid-loop-kaushik}
    \tcc{Use blocks of copy partition to stabilize blocks of original partition}
        \For{$B \in P$}{ \label{algo:inner-loop-kaushik}
            $\mathrm{succ} \gets B \cap N^{+}(B^{\cp})$\;\label{algo:kaushik-set-succ}
            $\mathrm{nonSucc} \gets B - N^{+}(B^{\cp})$\;
        
            \tcc{Split non-stable blocks}
            \If{ $\mathrm{succ}\neq \emptyset$ \KwAnd $\mathrm{nonSucc} \neq \emptyset$}{\label{algo:split-start}
                $P.\mathrm{add}(\mathrm{succ})$\;
                $P.\mathrm{add}(\mathrm{nonSucc})$\;
                $P.\mathrm{delete}(B)$\;
                $\mathrm{wasSplit} \gets \mathrm{true}$\;\label{algo:wasSplit-update}
            }\label{algo:split-end}
        }
        $\mathrm{usedSplitters}.\mathrm{add}(B^{\cp})$\;
    }\label{algo:mid-loop-kaushik-end}
    \If{$\neg\mathrm{wasSplit}$}{\label{algo:wasSplit-check}
        \KwBreak\; 
    }
}\label{algo:stabilizie-routine-end}
\Return $P$\;
}
\end{algorithm}


\subsection{Generic \BRS Algorithm}
\label{sec:brs}

The parallel \BRS algorithm is not specifically an implementation of $k$-bisimulation.
Rather, one can define a graph summary model in a formal language FLUID (see Section~\ref{sec:gsm}).
This model is denoted by~$\sim$ and input to the \BRS algorithm, which then summarizes a graph w.r.t.~$\sim$.
In particular, the $k$-bisimulation models of Schätzle \etal and Kaushik \etal can be expressed in FLUID, as shown in Section~\ref{sec:gsm}.
\todo{Response to Reviewer \#1 with some more details about the message passing.
See also comment further below.}
\response{Thus, the \BRS algorithm can compute $k$-bisimulation partitions.
In other words, $k$-bi\-sim\-ul\-ation can be incorporated into any graph summary model defined in FLUID.
The \BRS algorithm summarizes the graph w.r.t.~$\sim$ in parallel and uses the Signal/Collect paradigm.
In Signal/Collect~\cite{DBLP:journals/semweb/StutzSB16}, vertices collect information from their neighbors, sent over the edges as signals.
Details of the use of the Signal/Collect paradigm in our algorithm can be found in Blume \etal~\cite{DBLP:conf/cikm/BlumeRS20}.
Briefly, the algorithm builds equivalence classes by starting with every vertex in its own singleton set and forming unions of equivalent vertices.
Before outlining the algorithm, we give a necessary definition.

\begin{definition}
Suppose we have a graph summary $GS$ of $G=(V,E,l_V,l_E)$ w.r.t.\@ some GSM~$\sim$.
For each $v\in V$, the \emph{vertex summary} $vs$ is the subgraph of $GS$ that defines $v$'s equivalence class w.r.t.~$\sim$.
\end{definition}}

\extended{Our implementation of the \BRS algorithm is an adaptation of Blume \etal~\cite{DBLP:conf/cikm/BlumeRS20}.
Our version of the \BRS algorithm uses hash-based messages during the parallel computation of the graph summary and has been shown to be memory efficient~\cite{DBLP:journals/corr/abs-2111-12493-w-Jannik}. }%
We give the pseudocode of our version of the \BRS algorithm in Algorithm~\ref{alg:brs-adjusted} and briefly describe it below.
\cite{DBLP:journals/corr/abs-2111-12493-w-Jannik}~gives a step-by-step example run.

%
\begin{algorithm}[!h]
\small
\SetAlgoLined
\SetNoFillComment
\SetFuncSty{textsc}
\SetKwProg{Fn}{function}{}{end}
\SetKwProg{ForAllParallel}{for all}{ do in parallel}{end}
\SetKwProg{ForAll}{for all}{ do}{end}
\SetKwInput{Input}{Input}
\SetKw{Result}{returns}
\SetKwFunction{summarizeAdjusted}{\summarizeAdjustedMethod{}}
\SetKwFunction{extractVertex}{VertexSchema}
\SetKwFunction{extractEdge}{EdgeSchema}
\SetKwFunction{signalMessages}{SendMsgs}
\SetKwFunction{mergeMessages}{ReceiveMsg}
\SetKwFunction{index}{FindAndMerge}
\SetKwFunction{mergehash}{MergeAndHash}
\SetKwFunction{sendMessageToVertex}{SendMsg}
\caption{Parallel \BRS algorithm 
}
\label{alg:brs-adjusted}
\Fn{\summarizeAdjusted{$G, (\sim_s,\sim_p,\sim_o)^k$}}{
\Result{graph summary $SG$}
\BlankLine
\tcc{Initialization}
\ForAllParallel{$v \in V$}{\label{algo:initialisation}
    $v.id_{\sim_s} \gets \hash($\extractVertex{$v$, $G$, $\sim_s$, $\sim_p$}$)$\; \label{algo:initialise-id-subject}
    $v.id_{\sim_o} \gets \hash($\extractVertex{$v$, $G$, $\sim_o$, $\sim_p$}$)$\; \label{algo:initialise-id-object}
\label{algo:intialisation-end}}
\vspace{1ex}
\tcc{If $k = 1$, only signal edge labels and  $v.id_{\sim_o}$}
\If{k = 1}{\label{algo:signal-only-objId}
    \ForAllParallel{$v \in V$}{
        \ForAll{$w \in N^-(v)$}{
            \signalMessages{w, $\langle \ell(w,v), 0, v.id_{\sim_o} \rangle$}\; \label{algo:only-signal-t-o}
        }
        $M_o \gets \{\langle L, id_{\sim_o} \rangle \mid \langle L, id_{\sim_s},id_{\sim_o} \rangle \text{ was received}\}$\; \label{algo:merge-only-t-o}
        $v.id_{\sim_s} \gets v.id_{\sim_s} \oplus $ \mergehash{$M_o$}\; \label{algo:update-id-only-t-o}
    }\label{algo:signal-only-objId-end}}
\Else{\label{algo:signal-all}
    \tcc{Signal initial messages. Update $v.id_{\sim_s}$ and $v.id_{\sim_o}$}
    \ForAllParallel{$v \in V$}{\label{algo:initial-signal}
        \tcc{Message each in-neighbor}
        \ForAll{$w \in N^-(v)$}{
            \sendMessageToVertex{w, $\langle \ell(w,v), v.id_{\sim_s}, v.id_{\sim_o}\rangle$}\;
            \label{algo:signal-t-s-o-initial}
        } 
        \tcc{Collect all incoming messages of $v$}
        $M_s \gets \{\langle L,id_{\sim_s}\rangle \mid \langle L, id_{\sim_s}, id_{\sim_o} \rangle \text{ was received}\}$\; \label{algo:merge-t-s-initial} 
        $M_o \gets \{\langle L,id_{\sim_o}\rangle \mid \langle L, id_{\sim_s}, id_{\sim_o} \rangle \text{ was received}\}$\; \label{algo:merge-t-o} 
        \tcc{Update identifiers by hashing the messages}    
        $v.id_{\sim_s} \gets v.id_{\sim_s} \oplus \mergehash(M_s)$\; \label{algo:update-id-s-initial}
        $v.id_{\sim_o} \gets v.id_{\sim_o} \oplus \mergehash(M_o)$\; \label{algo:update-id-o-intitial}
    }\label{algo:initial-signal-end}
    
    \vspace{1ex}
   
    \tcc{Signal messages $k-2$ times. As above, but we do not include $L$ when updating $v.id_{\sim_o}$. (See text.)}
    \For{$i \gets 2$ \KwTo $k - 1$}{\label{algo:signal}
        \ForAllParallel{$v \in V$}{
            \ForAll{$w \in N^-(v)$}{
                \sendMessageToVertex{w, $\langle \ell(w,v), v.id_{\sim_s}, v.id_{\sim_o}\rangle$}\;
            }
            $M_s \gets \{\langle L,id_{\sim_s}\rangle \mid \langle L, id_{\sim_s}, id_{\sim_o} \rangle \text{ received}\}$\; \label{algo:merge-t-s} 
            $M_o \gets \{\langle id_{\sim_o} \rangle \mid \langle L, id_{\sim_s}, id_{\sim_o} \rangle \text{ received}\}$\; \label{algo:merge-o} 
            $v.id_{\sim_s} \gets v.id_{\sim_s} \oplus \mergehash(M_s)$\; \label{algo:update-id-s}
            $v.id_{\sim_o} \gets v.id_{\sim_o} \oplus  \mergehash(M_o)$\; \label{algo:update-id-o}
        }
}\label{algo:signal-end}
    \vspace{1ex}
    
    \tcc{Signal final messages. Update $v.id_{\sim_s}$}
    \ForAllParallel{$v \in V$}{ \label{algo:final-signal}
        \ForAll{$w \in N^-(v)$}{
            \sendMessageToVertex{w, $\langle \emptyset, 0, v.id_{\sim_o}\rangle$}\; \label{algo:signal-o}
        }
        $M_o \gets \{\langle id_{\sim_o} \rangle \mid \langle L, id_{\sim_s},id_{\sim_o} \rangle \text{ was received}\}$\; \label{algo:merge-o-final}
        $v.id_{\sim_s} \gets v.id_{\sim_s} \oplus$ \mergehash{$M_o$}\; \label{algo:update-id-s-final}
    } \label{algo:final-signal-end}
}\label{algo:signal-all-end}

\vspace{1ex}
$SG \gets$  \index{$SG, V$}\;\label{algo:adjusted-merge}
\Return $SG$\;
}
\end{algorithm}


\textit{Initialization (lines \ref{algo:initialisation}--\ref{algo:intialisation-end}).} 
For each vertex $v \in V$, \textsc{VertexSchema} computes the local schema information w.r.t.\@ $\sim_s$ and $\sim_o$ of the graph summary model $({\sim_s,} {\sim_p,} {\sim_o})^k$.
The method also takes into account $\sim_p$, \ie the equivalence relation defined over the edge labels.
Order-invariant hashes of this schema information are stored as the identifiers $id_{\sim_s}$ and $id_{\sim_o}$ (lines \ref{algo:initialise-id-subject}--\ref{algo:initialise-id-object}).

At the end of the initialization, every vertex has identifiers $id_{\sim_s}$ and $id_{\sim_o}$.
Two vertices $v$ and~$v'$ are equivalent w.r.t.\@ $(\sim_s, \sim_p, \sim_o)$, iff $v.id_{\sim_s} = v'.id_{\sim_s}$.
Thus, this initialization step can be seen as iteration $k=0$ of bisimulation.

\textit{Case of $k=1$ bisimulation
(lines \ref{algo:signal-only-objId}--\ref{algo:signal-only-objId-end}).} 
Every vertex~$v$ sends, to each in-neighbor~$w$, its $id_{\sim_o}$ value and the label set~$\ell(w,v)$ of the edge $(w,v)$  (line~\ref{algo:only-signal-t-o}).

Every vertex~$v$ sends, to each in-neighbor~$w$, its $id_{\sim_o}$ value and the label set \response{$L=\ell(w,v)$}\todo{Reviewer \#1: ``What is $L$?''} of the edge $(w,v)$  (line~\ref{algo:only-signal-t-o}).
Thus, each vertex receives a set of schema $\langle L, id_{\sim_o} \rangle$ pairs from its out-neighbors, which are collated into the set~$M_o$ (line~\ref{algo:merge-only-t-o}) and merged with an order-invariant hash to give $v$'s new $id_{\sim_s}$ (line~\ref{algo:update-id-only-t-o}).
\todo{Response to Reviewer \#1 explaining $\textsc{MergeAndHash}$ and $\oplus$.}
\response{
Here, the $\textsc{MergeAndHash}(M_o)$ function first merges the elements of the schema message $M_o$ received from vertex $o$ and hashes it with an order-independent hash function.
This hash is then combined with the existing hash value $v.id_{\sim_s}$ using the xor ($\oplus$) operator.
}

\textit{Case of $k>1$ bisimulation (lines \ref{algo:initial-signal}--\ref{algo:signal-all-end}).}
In the first iteration (lines \ref{algo:initial-signal}--\ref{algo:initial-signal-end}), every vertex $v$ sends a message to each of its out-neighbors~$w$. 
The message contains $v$'s $id_{\sim_s}$ and $id_{\sim_o}$ values, and the edge label set~$\ell(w,v)$ (line~\ref{algo:signal-t-s-o-initial}).
Subsequently, the incoming messages of the vertex are merged into a set of tuples with the received information $\langle \ell(w,v), id_{\sim_s} \rangle $ and  $\langle  \ell(w,v), id_{\sim_o} \rangle $ (lines \ref{algo:merge-t-s-initial} and~\ref{algo:merge-t-o}).
Finally, the identifiers $id_{\sim_s}$ and $id_{\sim_o}$ of $v$ are updated by hashing the corresponding set (lines \ref{algo:update-id-s-initial} and~\ref{algo:update-id-o-intitial}).
Note that, whenever an update of an identifier value $v.id_{\sim_s}$ of vertex $v$ is performed, the algorithm combines the old $v.id_{\sim_s}$ with the new hash value, indicated by~$\oplus$.

In the remaining iterations, the algorithm performs the same steps (lines \ref{algo:signal}--\ref{algo:signal-end}), but excludes the edge label set $\ell(w,v)$ when merging messages for $id_{\sim_o}$ (line~\ref{algo:merge-o}).
When merging the messages in $id_{\sim_o}$, it is not necessary to consider $\ell(w,v)$, as in the iterations $2$ to $k-1$ it is only needed to update the $id_{\sim_o}$ values using the hash function as described above. 
This is possible as $id_{\sim_o}$ by definition already contains the edge label set $\ell(w,v)$, computed in the first iteration.

In the final iteration, the identifiers are updated w.r.t.\@ the final messages (lines \ref{algo:final-signal}--\ref{algo:final-signal-end}).
The final messages received\extended{~by the vertices} are the values stored in the out-neighbors' $id_{\sim_o}$ values.
Each vertex signals its $id_{\sim_o}$ value to its in-neighbors (line~\ref{algo:signal-o}).
\shortorextended{The}{Again, the} messages a vertex receives are merged (line~\ref{algo:merge-o-final}) and hashed to update the final $id_{\sim_s}$ value (line~\ref{algo:update-id-s-final}).
Equivalence between any two vertices $v$ and~$v'$ can now be defined.
Vertices with the same $id_{\sim_s}$ value are merged (line~\ref{algo:adjusted-merge}), ending the computation.

\todo{Response to Reviewer \#2 on complexity of the algorithm.}
\response{We show in \cite{DBLP:journals/corr/abs-2111-12493-w-Jannik} that Algorithm~\ref{alg:brs-adjusted} computes $k$-bisimulation of a graph with $m$~edges in time $O(km)$.
As a modification of the algorithm in~\cite{DBLP:journals/tcs/BlumeRS21}, the algorithm is correct, as long as hash collisions are avoided.}

\section{Experimental Apparatus}
\label{sec:experiments}

\subsection{Datasets}
\extended{It is important to evaluate graph summarization algorithms on real\hyp{}world datasets, as synthetic datasets do not capture the characteristics of real\hyp{}world graphs~\cite{DBLP:conf/cikm/BlumeRS20,DBLP:conf/sigmod/LuoFHBW13}.
Thus, three real\hyp{}world datasets and two synthetic datasets were chosen for the experiments.}
We experiment with smaller and larger \shortorextended{as well as real-world and synthetic }{}graphs.
Table~\ref{tab:statistics-datasets} lists statistics of these datasets, where $r(l_V) = |\{l_V(v)\mid v\in V\}|$ is the number of different label sets (range) and $\mu(|l_V(v)|)$ is the average number of labels of a vertex $v \in V$.
\extended{
Further details are provided in Appendix~\ref{sec:appendix:datasets-stats}.
}

\begin{table}[t]
\small
\centering
\addtolength{\tabcolsep}{1ex}
        \begin{tabular}{|l|rrrrr@{$\,\pm\,$}lr|}
        \hline
        \multicolumn{1}{|c|}{Graph} & \multicolumn{1}{c}{$|V|$} & \multicolumn{1}{c}{$|E|$} & \multicolumn{1}{c}{$|\Sigma_V|$} & \multicolumn{1}{c}{$r(l_V)$} & \multicolumn{2}{c}{$\mu(|l_V(v)|)$} & 
         \multicolumn{1}{c|}{$|\Sigma_E|$} \\
        \hline
        Laundromat100M &  $30\,$M &   $88\,$M &  $33,431$ &   $7,373$ & $0.93$ & $44$ & $5,630$ \\
        BTC150M        &  $5\,$M &   $145\,$M &      $69$ &     $137$ & $1.04$ & $0.26$ & $10,750$ \\   
        BTC2B          & $80\,$M & $1.92\,$B & $113,365$ & $576,265$ & $0.95$ & $1.82$ & $38,136$ \\
        \hline
        BSBM100M       & $18\,$M &    $90\,$M &   $1,289$ &   $2,274$ & $1.02$ & $0.13$ & $39$ \\ 
        BSBM1B         & $172\,$M &  $941\,$M &   $6,153$ &  $27,306$ & $1.03$ & $0.18$ & $39$ \\ 
        \hline
        \end{tabular}
        %
        %
        %
        %
        %

\addtolength{\tabcolsep}{-1ex}
    \caption{Statistics of the \shortorextended{datasets}{real-world datasets (top) and synthetic datasets (bottom)}.}
    \label{tab:statistics-datasets}
\end{table}

\extended{\subsubsection{Real-World Datasets}}
Three real\hyp{}world datasets were chosen.
The \textit{Laundromat100M} dataset contains $100\,$M edges of the LOD Laundromat service~\cite{DBLP:conf/semweb/BeekRBWS14}.\extended{\footnote{\url{https://lodlaundromat.org} (offline, a local copy of the dataset was used).}}
This service automatically cleaned existing linked datasets and provided the cleaned version on a publicly accessible website.
\extended{The dataset consists of about $29.87$ million vertices and $87.85$ million edges.
It contains many more different labels ($33,431$) and label sets ($7,373$) than the other smaller datasets.
However, on average each vertex has only $0.93$ labels. }%
The \textit{BTC150M} and \textit{BTC2B} datasets contain, respectively, around $150$~million and $1.9$~billion edges of the Billion Triple Challenge 2019 (BTC2019)  dataset \cite{DBLP:conf/semweb/HerreraHK19}.\extended{\footnote{The full dataset has approximately $2.15$~billion edges. It is provided in several files. One file, containing about $150\,$M edges was corrupted and could not be parsed, so was omitted from our experiments.} Herrera, Hogan and K\"afer statistically analyzed the entire BTC2019 dataset.
They found that} 93\% of the total edges \extended{(${\sim}2$~billion)~}originate from Wikidata~\cite{wikidata}.
BTC150M is the first chunk of the $1.9$~billion edges.
\extended{Moreover, investigation of used vocabularies, predicates, and classes, led Herrera \etal to the conclusion that BTC2019 is a ``highly diverse dataset'' \cite{DBLP:conf/semweb/HerreraHK19}.
The BTC150M dataset has about $5$~million vertices and $145$~million edges.
BTC150M contains the fewest different labels ($69$) and label sets ($137$) of the smaller datasets.
However, on average a vertex has $12.38$ different labels, about $12$ times more than the other two.
Also, BTC150M has the most different edge labels ($10,750$).
BTC2B has about $79.65$ million vertices, $1.92$ billion edges, $113,365$ different label sets and $38,136$ different edge labels.
On average, a vertex has $10.40$\extended{~different} labels, the most of the larger datasets.}
\extended{

\subsubsection{Synthetic Datasets}}
For synthetic datasets, two versions of the Berlin SPARQL Benchmark (BSBM) \cite{DBLP:journals/ijswis/BizerS09} were used.
\extended{Originally developed for ``comparing the SPAQRL query performance of native RDF stores with that of SPARQL-to-SQL rewriters''~\cite{DBLP:journals/ijswis/BizerS09}, the benchmark consists of a data generator, used here, and a test driver.} The BSBM data generator produces RDF datasets \shortorextended{that simulate}{which consist of products, vendors, offers, and reviews, simulating} an e-commerce use case.
\extended{Datasets of arbitrary sizes can be generated by specifying the numberof products represented.} \textit{BSBM100M} was generated with $284,826$ products and has about $17.77$~million vertices and $89.54$~million edges. \extended{ It has the fewest edge labels of the smaller datasets: on average~$39$. } 
\textit{BSBM1B} was generated with $2,850,000$ products and has about $172$~million vertices and $941$~million edges.
\extended{ It has $6,153$ different label sets and $39$ different edge labels.
Other synthetic benchmark datasets such as LUBM\footnote{\url{http://swat.cse.lehigh.edu/projects/lubm/}} and  SP$^2$Bench\footnote{\url{http://dbis.informatik.uni-freiburg.de/index.php?project=SP2B}} could be used in further experiments.}

\subsection{Procedure}

An experiment consists of the algorithm to run, the dataset to summarize, and the bisimulation degree~$k$.
In case of the \BRS algorithm, it additionally consists of the graph summary model to use.
We have two different graph summary models defined by Schätzle \etal~\cite{DBLP:conf/sigmod/SchatzleNLP13} and Kaushik \etal \cite{DBLP:conf/icde/KaushikSBG02}.
Each model comes in two implementations, one native implementation as defined by the original authors and a generic implementation through our hash-based version of the \BRS algorithm. 
\todo{Response to Review \#1 on the motivation to choose Kaushik \etal and Schätzle \etal}
\response{We chose the bisimulation algorithms of Kaushik \etal and Schätzle \etal as they represent two typical variants of backward and forward $k$-bisimulation models as found in the literature (cf.\@ Section~\ref{sec:related_work}).
This choice of algorithms allows us to demonstrate that our algorithm can be applied to different settings.}
We use the terms \emph{\BRS-Sch\"atzle} and \emph{\BRS-Kaushik} to refer to our implementations of the two GSMs \BRS algorithm; we refer to our single-purpose implementations of these two GSMs as \emph{native Sch\"atzle} and \emph{native Kaushik}.

The four \extended{implemented }algorithms are applied on \extended{each of }the five datasets, \shortorextended{giving}{for a total of} 20 experiments. 
Each experiment is executed with a bisimulation degree of $k=1, \dots, 10$, using the following procedure.
We run the algorithms six times\extended{ in a row} with the specific configuration.
We use the first run\extended{~(run 0)} as a warm-up\extended{~run} and\extended{~hence} do not account it for our measurements\extended{~of the dependent variables}. %
The next five runs\extended{~(runs 1--5)} are used to measure the variables.
\extended{Thus, the last five runs of all experiments follow the same environment, \ie none of the experimental results are influenced by any processes that ran before the experiment started.}

\subsection{Implementation}
\label{sec:implementation}

\todo{Response to Reviewer \#1 with some more details about the message passing.
See also comment above.}
\response{All algorithms, \ie the native algorithms of Schätzle \etal and Kaushik \etal 
and their generic \BRS-variants are implemented using the same underlying framework and paralellization approach, \ie are implemented in Scala upon the Apache Spark Framework.
This API offers flexible support for parallel computation and message passing, which enables implementation of Map-Reduce and Signal-Collect routines.}
\extended{For the sequential native Kaushik algorithm (Algorithm~\ref{alg:kaushik-alg}), the Apache Spark API was used to implement the graph data structure, the partition data structures, and to parse and initialize these data structures. %
The stabilization routine of the native Kaushik algorithm (lines \ref{algo:stabilizie-routine-start}--\ref{algo:stabilizie-routine-end}) is implemented sequentially and has limited scope for parallelization.
The first inner loop (line~\ref{algo:mid-loop-kaushik}) has the following problem:
In each iteration of the second inner loop at line~\ref{algo:inner-loop-kaushik}, there is a chance that new blocks are added to the current partition~$P_i$.
This leads to two new blocks which must be checked for instability with respect to all the other block-copies in the copy partition.} %
\shortorextended{We}{For our experiments, we} use an Ubuntu 20 system with 32~cores and $2\,\mathrm{TB}$ RAM\extended{, with exclusive access to avoid interference from other processes}.
The Apache Spark contexts were given the full resources\extended{, \ie 32 cores and $1.94\,\mathrm{TB}$ heap space}.
Time and memory measurements were taken using the Apache Spark Monitoring API.

\subsection{Measures}
\label{sec:metrics}
We evaluate the algorithms' running time and memory consumption.
For every run of an experiment, we report the 
total run time,
the run time of each of the $k$ iterations, and
the maximum JVM on-heap memory \shortorextended{consumption.}{consumption during execution.}

\section{Results}
\label{sec:results}

\todo{Made clearer that Figure 2 contains the main results as it reports the runtimes per iteration (each value of~$k$).}
\response{We present full results for each algorithm, iteratively calculating $k$-bisimulation for every value of $k=1,\dots, 10$ and every dataset, in Figure~\ref{fig:results-smaller}.
Table~\ref{tab:results-runtime} summarizes the average total run time (minutes) for each experiment, for the computation of $10$-bisimulation.}
The \BRS algorithm takes an additional initialization step (see Algorithm~\ref{alg:brs-adjusted}, lines \ref{algo:initial-signal}--\ref{algo:initial-signal-end}).
\extended{A breakdown can be found in Appendix~\ref{sec:appendix:breakdown-brs-with-initialization}. }%
Table~\ref{tab:results-memory} reports the maximum JVM on-heap memory in GB\extended{~used during the execution} for each experiment.
\shortorextended{%
The \BRS-Schätzle algorithm computes the $10$-bisimulation the fastest on all datasets, except for BSBM100M, where \BRS-Kaushik is fastest.
Native Schätzle consumes the least memory on all smaller datasets.
Native Schätzle consumes slightly more memory on BSBM1B than \BRS, whereas on BTC2B, the memory consumption is about the same.
}
{In summary, on the smaller datasets the \BRS-Schätzle algorithm computes the $10$-bisimulation the fastest on BTC150M and Laundromat100M, taking $4.08$ and $5.56$ minutes on average.
On BSBM100M, \BRS-Kaushik runs the fastest, taking $5.20$ minutes on average.
Native Schätzle consumes the least memory on all smaller datasets.
Concerning the larger datasets BTC2B and BSBM1B, \BRS-Schätzle computes the ten iterations the fastest taking $61.96$ and $54.44$ minutes on average respectively.
Native Schätzle consumes slightly more memory than \BRS-Schätzle on BSBM1B, whereas on BTC2B \BRS and native Schätzle consume about the same amount of memory.}

\newcommand{\figscale}{0.382\linewidth} 
\extended{
\todoyellow{Has to be set for the larger version:}
 Large: 0.45 
}

\begin{figure*}[h!]
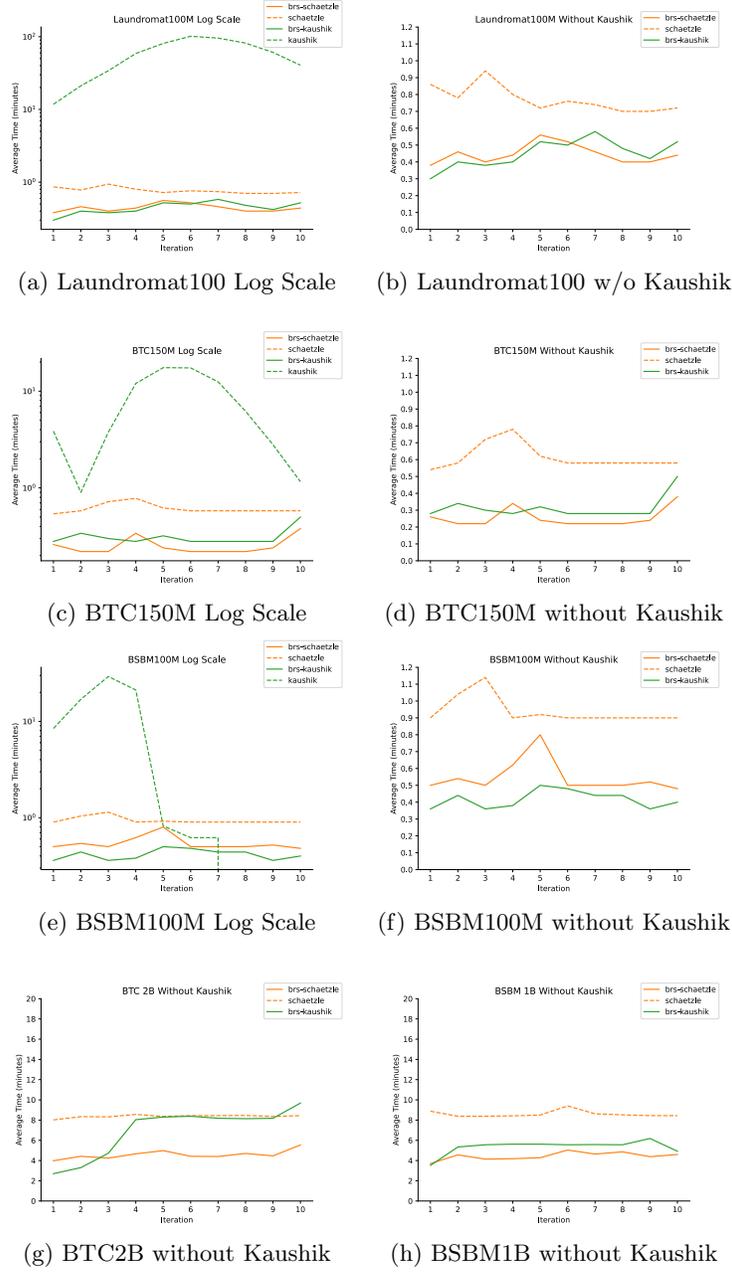
%
    \centering
    \subfloat[Laundromat100 Log Scale ]{
        \includesvg[width=\figscale]{images/plots/iterationTimes_laundromat100M_log.svg}
        \label{plt:laundromat100-log}
    }
    \subfloat[Laundromat100 w/o Kaushik ]{
        \includesvg[width=\figscale]{images/plots/iterationTimes_laundromat100M_noKaushik.svg}
        \label{plt:laundromat100-noKaushik}
    }
    \\
    \subfloat[BTC150M Log Scale]{
        \includesvg[width=\figscale]{images/plots/iterationTimes_btc150M_log.svg}
        \label{plt:btc150M-log}
    }
    \subfloat[BTC150M without Kaushik]{
        \includesvg[width=\figscale]{images/plots/iterationTimes_btc150M_noKaushik.svg}
        \label{plt:btc150M-noKaushik}
    }
    \\
    \vspace{-3mm}
    \subfloat[BSBM100M Log Scale]{
        \includesvg[width=\figscale]{images/plots/iterationTimes_bsbm100M_log.svg}
        \label{plt:bsbm100M-log}
    }
    \subfloat[BSBM100M without Kaushik]{
        \includesvg[width=\figscale]{images/plots/iterationTimes_bsbm100M_noKaushik.svg}
        \label{plt:bsbm100M-noKaushik}
    }
\shortorextended{
    \\
    \subfloat[BTC2B without Kaushik]{
        \includesvg[width=\figscale]{images/plots/iterationTimes_btc2b_noKaushik.svg}
        \label{plt:btc2b}
    }
    \subfloat[BSBM1B without Kaushik]{
        \includesvg[width=\figscale]{images/plots/iterationTimes_bsbm1B_noKaushik.svg}
        \label{plt:bsbm1b}
    }
}{}
    \caption{Average iteration times (minutes) \extended{of experiments~}on the smaller datasets\shortorextended{~in (a) to (f) and larger datasets in (g) and (h).}{}
    }
    \label{fig:results-smaller}
\end{figure*}

\newcommand{\mytextsubscript}[1]{{\color{black}\textsubscript{~(#1)}}}

\begin{table}[!t]
\small
\centering
\addtolength{\tabcolsep}{1ex}
        \begin{tabular}{|l|r|r|r|r|}
        \hline
         & \multicolumn{2}{c|}{Schätzle \etal} & \multicolumn{2}{c|}{Kaushik \etal}  \\
         \cline{2-5}
         & \multicolumn{1}{c|}{\BRS} & \multicolumn{1}{c|}{Native} & \multicolumn{1}{c|}{\BRS} & \multicolumn{1}{c|}{Native} \\
         \hline
            Laundromat100M & 
            5.56\mytextsubscript{0.50} & 
               7.72\mytextsubscript{0.07} & 
               5.60\mytextsubscript{0.15} & 
               586.66\mytextsubscript{21.09} \\
            BTC150M & 
              4.08\mytextsubscript{1.55} & 
              6.14\mytextsubscript{0.38} & 
              4.54\mytextsubscript{1.34} & 
              78.02\mytextsubscript{3.71} \\
            BTC2B & 
              61.96\mytextsubscript{11.38} & 
              83.74\mytextsubscript{1.96} & 
              85.92\mytextsubscript{13.6} & 
              out of time \\
            \hline
            BSBM100M & 
              6.46\mytextsubscript{0.08} & 
              9.40\mytextsubscript{0.06} & 
              5.20\mytextsubscript{0.50} & 
              77.84\mytextsubscript{2.41} \\
            BSBM1B & 
              54.44\mytextsubscript{4.35} & 
              85.98\mytextsubscript{3.83} & 
              64.00\mytextsubscript{3.22} & out of memory \\
        \hline
        \end{tabular}
\addtolength{\tabcolsep}{-1ex}
        \caption{Total run time (in minutes) needed for computing the $k=1,\dots,10$ bisimulation (average and standard deviation over 5 runs). 
        \extended{Top are real-world datasets, bottom are synthetic datasets. 
        \BRS and Native Schätzle \etal are parallel algorithms, while Native Kaushik \etal is sequential.}}
    \label{tab:results-runtime}

\end{table}

\begin{table}[!h]
\small
    \centering
\addtolength{\tabcolsep}{1ex}
    \begin{tabular}{|l|r|r|r|r|}
    \hline
         & \multicolumn{2}{c|}{Schätzle \etal} & \multicolumn{2}{c|}{Kaushik \etal}  \\
         \cline{2-5}
             & \multicolumn{1}{c|}{\BRS} & \multicolumn{1}{c|}{Native} & \multicolumn{1}{c|}{\BRS} & \multicolumn{1}{c|}{Native} \\
        \hline
        Laundromat100M & 211.5 & 147.9 & 210.5 & 335.1 \\
        BTC150M & 140.6 & 107.7 & 130.1 & 181.7 \\
        BTC2B & 1,249.1 & 1,249.7 & 1,249.3 & out of time \\
        \hline
        BSBM100M & 248.6 & 113.1 & 172.0  & 327.0 \\
        BSBM1B & 1,248.2 & 1,335.4 & 1,249.2 & out of memory \\
    \hline
    \end{tabular}
\addtolength{\tabcolsep}{-1ex}
    \caption{Maximal JVM on-heap memory (in GB) used for $k=1,\dots,10$ bisimulation.
    We provide the maximal memory usage determined over five runs, instead of the average and standard deviation, in order to demonstrate what memory is needed to perform the bisimulations without running out of memory. 
    }
    \label{tab:results-memory}
\end{table}

%

\textit{\textbf{Smaller Datasets (100M+ Edges)}.}
Figure~\ref{fig:results-smaller} shows the average run time (minutes) for each of the ten iterations on the smaller datasets.
The native Kaushik experiments take much longer than the others (Figures~\ref{plt:laundromat100-log}, \ref{plt:btc150M-log}, and~\ref{plt:bsbm100M-log}), so we provide plots without native Kaushik (Figures~\ref{plt:laundromat100-noKaushik}, \ref{plt:btc150M-noKaushik}, and~\ref{plt:bsbm100M-noKaushik}), to allow easier comparison between the other algorithms.

\shortorextended{\BRS-Schätzle and native Schätzle have relatively constant iteration time on all smaller datasets.}{\BRS-Schätzle has relatively constant iteration time on all smaller datasets. On Laundromat100M (Figure~\ref{plt:laundromat100-noKaushik}), it computes an iteration in about $0.4$ to $0.6$ minutes.
On BTC150M (Figure~\ref{plt:btc150M-noKaushik}) an iteration takes approximately $0.2$ to $0.4$ minutes and on BSBM100M (Figure~\ref{plt:bsbm100M-noKaushik}) $0.5$ to $0.8$ minutes.
The total run time (Table~\ref{tab:results-runtime}) is lowest on BTC150M, taking about $4.08$ minutes on average.
On Laundromat100M, the average run time is about $5.56$ minutes and on BSBM100M, the algorithm needs $6.46$ minutes on average.
Native Schätzle shows a similar behavior, with a relatively constant iteration time across the datasets.}
\shortorextended{For example, on}{On} 
Laundromat100M (Figure~\ref{plt:laundromat100-noKaushik}), native Schätzle computes an iteration in about $0.7$ to $0.95$ minutes.
\extended{On BTC150M (Figure~\ref{plt:btc150M-noKaushik}) an iteration takes approximately $0.55$ to $0.8$ minutes.
Finally, on BSBM100M (Figure~\ref{plt:bsbm100M-noKaushik}) the average iteration time ranges from $0.9$ to $1.15$ minutes.
The average run time (Table~\ref{tab:results-runtime}) is fastest on BTC150M with $6.14$ minutes on average.
This is followed by an average run time of $7.72$ minutes on Laundromat100M and $9.40$ minutes on BSBM100M. }%
\BRS is slightly faster on all smaller datasets, but uses more memory (Table~\ref{tab:results-memory}).
\BRS-Kaushik shows similar results to \BRS-Schätzle.
Again iteration time is relatively constant on all datasets\extended{~taking about $0.3$ to $0.6$ minutes on Laundromat100M (Figure~\ref{plt:laundromat100-noKaushik}) and $0.3$ to $0.5$ minutes on BTC150M and BSBM100M (Figures \ref{plt:btc150M-noKaushik} and~\ref{plt:bsbm100M-noKaushik})}. 
As before, it is fastest on BTC150M\extended{~with, an average run time of $4.54$ minutes}.
\extended{On BSBM100M and Laundromat100M, the average run times are about $5.20$ and $5.60$ minutes, respectively.}%

Native Kaushik shows a different behavior across the datasets.
Iteration time varies on all datasets: the iteration time increases in early iterations until it reaches a maximum value, and then decreases.
The only exception to this behavior occurs on BTC150M, where the iteration time decreases from iteration one to two before following the described pattern for the remaining iterations.
\shortorextended{For example, on}{On} 
Laundromat100M (Figure~\ref{plt:laundromat100-log}) an iteration takes about $10$ to $100$ minutes.
The maximum iteration time is reached in iteration six.
\extended{On BTC150M (Figure~\ref{plt:btc150M-log}) the iteration time ranges from approximately $1$ to $17$ minutes.
Here, the maximum iteration time is reached in iteration five.
Finally, on BSBM100M (Figure~\ref{plt:bsbm100M-log}) the algorithm needs about $0.6$ to $30$ minutes to compute an iteration.
The maximum iteration time is iteration three on this dataset.
Furthermore, the full bisimulation w.r.t.\@ Definition~\ref{def:vertex-labeled-bw-bisimulation} is reached and detected by the algorithm at iteration seven. }%
Native Kaushik runs fastest on BSBM100M, taking about $77.84$ minutes on average.
\extended{This is followed by an average run time of $78.02$ minutes on BTC150M and $586.66$ minutes on Laundromat100M. }%
\BRS-Kaushik is much faster on all smaller datasets (Table~\ref{tab:results-runtime}) and uses slightly less memory (Table~\ref{tab:results-memory}).

\textit{\textbf{Larger Datasets (1B+ Edges)}.}%
\extended{
\begin{figure}%
    \centering
    \subfloat[BTC2B without Kaushik \etal]{
        \includesvg[width=0.8\linewidth]{images/plots/iterationTimes_btc2b_noKaushik.svg}
        \label{plt:btc2b}
    }
    \\
    \subfloat[BSBM1B without Kaushik \etal]{
        \includesvg[width=0.8\linewidth]{images/plots/iterationTimes_bsbm1B_noKaushik.svg}
        \label{plt:bsbm1b}
    }
    \caption{Average iteration times (minutes) of experiments on the larger datasets.}
    \label{fig:results-larger}
\end{figure}

Figure~\ref{fig:results-larger} shows plots of the average run time (minutes) for each of the ten iterations on the larger datasets.}
\BRS-Schätzle computes the $k=1, \dots, 10$ bisimulation on BTC2B and BSBM1B with a relatively constant iteration time.
\extended{On BTC2B (Figure~\ref{plt:btc2b}) an iteration takes approximately $4$ to $6$ minutes and on BSBM100M (Figure~\ref{plt:bsbm1b}) $4$ to $5$ minutes. }%
The total run time (\shortorextended{Table~\ref{tab:results-runtime}}{Table~\ref{tab:results-runtime,tab:brs-init-iter-runtime}}), is lowest on BSBM1B: $54.44$ minutes on average.
\extended{On BTC2B the \BRS-Schätzle algorithm needs $61.96$ minutes on average per iteration. }%
Native Sch\"atzle also has nearly constant running time across iterations.
\extended{On BTC2B (Figure~\ref{plt:btc2b}) an iteration takes approximately $8$ minutes.
On BSBM1B (Figure~\ref{plt:bsbm100M-noKaushik}) the average iteration time ranges from $8$ to $9.5$ minutes. }%
The average running time for all ten iterations is lowest on BTC2B: $83.74$ minutes on average, compared to $85.98$ minutes on BSBM1B.
Comparing the run times of the two implementations, \BRS is $37\%$ on the BSBM1B dataset and $26\%$ faster on BTC2B.
Both algorithms require about the same memory during execution on BTC2B.
On BSBM1B, native Schätzle uses slightly more memory than \BRS.
Native Kaushik did not complete one iteration on BTC2B in $24$ hours, so execution was canceled.
The algorithm ran out of memory on BSBM1B (Table~\ref{tab:results-memory}).
\extended{Hence, the plots in Figure~\ref{fig:results-larger} only provide results for \BRS and native Schätzle.}

Finally, \BRS-Kaushik also has similar iteration times.
On BSBM1B (Figure~\ref{plt:bsbm1b}), computing a bisimulation iteration ranges from about $4$ to $6$ minutes.
On BTC2B (Figure~\ref{plt:btc2b}), the execution times for iterations one to three range from $2$ to $4$ minutes.
Iterations four to nine take about $8$~minutes each.
For the final iteration ten, the running time slightly increases to about $9.5$ minutes.

\section{Discussion}
\label{sec:discussion}

\noindent\textbf{\textit{Main Results}.}
\todo{Rewrote main results to make the contribution clearer (response to Reviewer \#1 and \#2).}
\response{Our results show that, on all datasets, the generic \BRS algorithm outperforms the native bisimulation algorithms of Sch\"atzle \etal and Kaushik \etal for $k=10$ (see Table~\ref{tab:results-runtime}).
Since the \BRS algorithm has an initialization phase, which is not present in the two native algorithms, we examine the data more closely to see at which value of~$k$ \BRS begins to outperform the native implementations.
We provide the exact numbers per iteration together with the standard deviation and averaged over five executions in~
Appendix~\ref{sec:appendix-results}.
%
Our generic \BRS-Kaushik outperforms the native Kaushik \etal for all~$k$: this is unsurprising, as \BRS parallelizes the original sequential algorithm.
The time taken to compute $k$-bisimulation is the total time for iterations up to, and including,~$k$.
The comparison of \BRS-Schätzle with the native Schätzle \etal algorithm shows that our generic algorithm begins to outperform the native one at $k=3$ for the Laundromat100M, BSBM100M, and BSBM1B; at $k=4$ for BTC150M; and at $k=5$ for BTC2B.
Again, we emphasize that \BRS is a generic algorithm supporting all graph summary models definable in FLUID (see introduction), whereas native algorithms compute only one version of bisimulation each.
}

\extended{The parallel algorithms \BRS-Schätzle, \BRS-Kaushik, and native Schätzle compute all of these ten bisimulation iterations, taking in total  about $4.08$ to $9.40$ minutes on the smaller datasets.
%
Furthermore, both parallel algorithms perform much faster than the sequential Kaushik \etal algorithm.
\BRS takes $4.08$ to $5.60$ and native Schätzle $6.14$ to $9.40$ minutes on the smaller datasets, whereas the sequential algorithm takes $77.84$ to $586.66$ minutes.
Hence, even for smaller graphs of $100M$ to $150M$ triples, we observe that our parallel algorithm is much faster than the sequential alternative. 
The parallel algorithms perform well on the synthetic BSBM1B graph and the real-world BTC2B graph.
The sequential native Kaushik algorithm takes one to two orders of magnitude longer and was unable to compute a $10$-bisimulation at all for the two largest datasets.}

\todo{Additional response to Reviewer \#2 to reflect on the choice of $k$, reaching full bisimulation, etc.}
\response{Note that, for each dataset and each algorithm, we iteratively compute in total ten bisimulation-based graph summaries using every value of~$k$ from $1$ to~$10$. 
As can be seen in Figure\extended{s}~\ref{fig:results-smaller}\extended{ and~\ref{fig:results-larger}}, the execution times per bisimulation iteration are fairly constant over all iterations for all datasets
for \BRS and Sch\"atzle \etal
The native Kaushik \etal algorithm sequentially computes the refinement of the partitions, so execution time directly relates to the number of splits per $k$-bisimulation iteration.
This can be seen by the curve in the plots of the smaller datasets (see
Figures~\ref{plt:laundromat100-log}, \ref{plt:btc150M-log}, and \ref{plt:bsbm100M-log}).
In particular, Kaushik \etal detect that full bisimulation has been reached and do not perform further computations: this happens after computing the 7th iteration ($7$-bisimulation, see Figure~\ref{plt:bsbm100M-log}) on the BSBM100M dataset but not on the other datasets on which the execution of the Kaushik \etal algorithm completed.
This kind of termination check could also be added to the \BRS algorithm but this is nontrivial, as we discuss below.}

\response{This consistent per-$k$ iteration runtime is explained by each iteration of the \BRS algorithm processing every vertex, considering all its neighbors.
Thus, the running time of the iteration depends primarily on the size of the graph, and not on how much the $k$-bisimulation that is being computed differs from the $(k-1)$-bisimulation that was computed at the previous iteration.
The native Schätzle \etal algorithm behaves in the same way.
There is some variation in per-iteration execution time but this may be because, as each vertex is processed, duplicate incoming messages must be removed.
The time taken for this will depend on the distribution of the incoming messages, which will vary between iterations.}

\response{We consider experimenting with values between $k=1$ and $10$ to be appropriate.
The native Kaushik \etal algorithm terminates when full bisimulation is reached, which shows that full bisimulation is reached at $k=7$ on the BSBM100M dataset, but has not been reached up to $k=10$ for Laundromat100 and BTC150M.
The rapidly decreasing per-iteration running times for  native Kaushik \etal for BTC150M (Figure~\ref{plt:btc150M-log}) suggests that full bisimulation will be reached in a few more iterations.
A similar curve can be observed for Laundromat100 (Figure~\ref{plt:laundromat100-log}), but the running time for the $k=10$ iteration is higher, suggesting that full bisimulation will not be reached for several iterations.
Thus, on these datasets, computing $10$-bisimulations is a reasonable thing to do, as full bisimulation has not yet been reached.
Note that, for the other algorithms that we consider (native Schätzle, and the two instances of \BRS), per-iteration running time is largely independent of~$k$ and of whether full bisimulation as been reached (see the per-iteration measurements on BSBM100M in 
Table~\ref{tab:detailed-results-bsbm100m}). 

Conversely, our primary motivation is graph summarization.
When two vertices in a graph are $10$-bisimilar, this means they have equivalent neighborhoods out to distance~$10$.
This means that they are already ``largely similar'' for that value of~$k$.
Thus, we feel that, having computed $k$-bisimulation for a relatively large value of~$k$, there is little advantage in going to $k+1$.
In other words, adding another iteration of $k+1$ still leaves us with vertices that are ``largely similar''.}

\response{
\textit{\textbf{Scalability to Graph Size}.}\extended{The \BRS algorithm computes $k$-bisimulation for a graph with $m$~edges in time $O(km)$~\cite{DBLP:journals/corr/abs-2111-12493-w-Jannik}.  This is linear in the size of the graph for fixed~$k$. 
Sch\"atzle \etal~\cite{DBLP:conf/sigmod/SchatzleNLP13} report an approximately linear empirical running time for their algorithm.
This cannot be compared directly with our work, as they are computing full bisimulations; however, they are using highly structured synthetic datasets in which full bisimulation is reached after a fixed number of iterations regardless of the dataset's size.
Thus, we would expect the Schätzle \etal algorithm to run in approximately linear time for $k$-bisimulation with $k$~fixed.
Kaushik \etal also report a running time of $O(km)$ for their algorithm~\cite{DBLP:conf/icde/KaushikSBG02}.} 
From the execution time of computing $k$-bisimulation for $k=1, \dots, 10$, we observe that \BRS-Schätzle, \BRS-Kau\-shik, and native Schätzle scale linearly.
Here, we consider the scalability of the algorithms with respect to graph size.
To this end, we fix on iterations $k=1,\dots, 10$ and compare the total time taken to compute these bisimulations for input graphs of different (large) sizes.
BTC150M contains approximately $5$M vertices and $145$M edges.
BTC2B has around $80$M vertices and $2$B edges, which is a factor of about $18$ and $14$ larger, respectively.
\BRS-Schätzle takes $4.08$ minutes, 
\BRS-Kaushik $4.54$ minutes, and native Schätzle $6.14$ minutes on BTC150M.
On BTC2B, they take $15$ times, $19$ times, and $14$ times longer.
BSBM1B contains about $10$ times as many vertices and edges as BSBM100M.
The experiments on BSBM1B took about $10$ times longer for \BRS-Schätzle, $12$ times for \BRS-Kaushik, and $9$ times for native Schätzle.
Thus, the scaling factor of the execution times is approximately equal to that of the graph's size.
Finally, the total runtimes of native Kaushik on the different graphs indicate that the algorithm does not scale linearly with the input graph's size.
The initial partition~$P_0$ has one block per label set and Laundromat100M has many different label sets (Definition~\ref{def:vertex-labeled-bw-bisimulation}).
The algorithm checks stability of every block against every other, leading to a run time that is quadratic in the number of blocks.
\extended{Details are in Appendix~\ref{sec:app:detailed-discussion-kaushik}.}
}

\textit{\textbf{Generalization and Threats to Validity}.}%
\label{sec:threat-to-validity}
We use synthetic and real-world graphs, which is important to analyze the practical application of an algorithm~\cite{DBLP:conf/cikm/BlumeRS20,DBLP:conf/sigmod/LuoFHBW13}. 
Two GSMs were used for evaluation of our hash-based \BRS algorithm.
The GSM of Schätzle \etal computes a forward $k$-bisimulation, based on edge labels (Definition~\ref{def:edge-labeled-fw-bisimulation}).
The GSM of Kaushik \etal computes a backward $k$-bisimulation based on vertex labels (Definition~\ref{def:vertex-labeled-bw-bisimulation}).
Hence, the two GSMs consider different structural features for determining vertex equivalence.
\response{Regardless, for both GSMs, the \BRS algorithm scales linearly with the number of bisimulation iterations and the
input graph's size and computes the aggregated $k=1,\dots,10$-bisimulations the fastest on every dataset.}

All algorithms are implemented in Scala, in the same framework.
\extended{Optimizations from the Spark Tuning Guide~\cite{tuning-spark}
were used equally to implement the algorithms such as arrays of objects (tuples) and primitives (integers). }%
Each algorithm was executed using the same procedure on the same machine with exclusive access during the experiments.
\extended{Hence, the experiments for each of the algorithms were performed in the same environment. }%
Each experimental configuration was run six times\extended{, to account for possible side effects from the hardware caching data and other machine properties}.
The first run is discarded in the evaluations to address\extended{ potential} side effects\extended{, leaving the reported averages over five independent measurements}.

\extended{Our implementation of the native Schätzle \etal algorithm follows the original work~\cite{DBLP:conf/sigmod/SchatzleNLP13}, using sets of tuples for sending and merging messages.
This could be optimized by using arrays of tuples instead.
Another optimization could be to remove the check at each iteration whether a full bisimulation has been reached (line~\ref{algo:check-count} in Algorithm~\ref{alg:schaetzle-alg}).}
\extended{To do this, Schätzle \etal count the distinct block identifiers of the current iteration.
They evaluated their algorithm experimentally on different versions of three synthetic benchmarks and on one real world graph~\cite{DBLP:conf/sigmod/SchatzleNLP13}.
On the synthetic datasets, full bisimulation was reached in iterations four (SP$^2$Bench), seven (LUBM), and twelve to fourteen (SIB).
The real-world graph (DBPedia version $3.7$) reached full bisimulation at iteration $37$.
Hence, together with the results provided here, where no graph reaches full bisimulation w.r.t.\@ Definition~\ref{def:edge-labeled-fw-bisimulation} in the first $10$ iterations, this step could be considered as unnecessary computing for the stratified bisimulations.
However, we implement the algorithms as originally defined; optimizations are left for future work.}

\extended{\textit{Future Work.} The provided results can be used as a foundation for several future works.
First, it would be interesting to see the performance of the \BRS algorithm for more GSMs that can be found in the literature.
Blume \etal~\cite{DBLP:journals/tcs/BlumeRS21} provide a large overview of existing GSMs that could be evaluated for values of~$k$.

Further, the concept of operating on blocks during execution, rather than on vertices and edges, which is applied by the sequential Kaushik \etal algorithm, could be evaluated for a parallel implementation.
Rajasekaran and Lee~\cite{DBLP:journals/tpds/RajasekaranL98} provide a parallel version of Paige and Tarjan's \cite{DBLP:journals/siamcomp/PaigeT87} optimized algorithm for computing relational coarsest partition problems (Definition~\ref{def:rel-coar-par-prob}).
However, one would have to adapt the implementation to a stratified bisimulation, as the relational coarsest partition corresponds to a full bisimulation \cite{DBLP:journals/iandc/KanellakisS90}.
Alternatively, Algorithm~\ref{alg:kaushik-alg} could be adjusted as described below:
A possible improvement could be to compute both loops in parallel and after the first inner loop is finished for all the block-copies $B^{\cp}$ in the copy partition, the newly created blocks are exchanged between all of them. 
This is repeated until no blocks are created anymore.
To implement this, a worker which is for example assigned to stabilize block $B_{i1}$ with respect to block-copy $B^{\cp}_{i5}$ needs to know, or need to be able to compute the neighbors of the vertices in $B^{\cp}_{i5}$.
This either results in more communication (\eg send the neighbors to the worker who is assigned for the respective block-copy) or in more memory consumption (e.g. broadcast a read-only map containing (vertex $\rightarrow$ list of neighbors)-entries of all vertices to every worker).

It would be very interesting to extend this work to incremental computation of $k$-bisimulation on evolving data graphs.
Doing this within a small memory footprint is a huge challenge~\cite{DBLP:journals/corr/abs-2111-12493-w-Jannik}.
The algorithm is complex, as a change in the input graph can change a vertex's id values in each iteration $i \in \{1,2,\ldots,k\}$.
Moreover, changes in one iteration can affect all subsequent iterations, ultimately, changing the final relationship between any two vertices.

We also experimented with a larger \extended{graph with $14\,\mathrm{B}$ triples.
We created a~}Laundromat14B dataset, which contains about $1.46$ billion vertices and $13.52$ billion edges.
\extended{The vertices in the graph exhibit $0.71$ million different label sets and the graph has $1.02$ million edge labels.}
None of the algorithms considered could compute $10$-bisimulation within the $2\,\mathrm{TB}$ of memory available.
Our focus is on comparing the algorithms' performance, so working in main memory is reasonable to eliminate effects such as network latency\extended{~in a distributed setting} or disk access times.
However, this motivates the need for external memory solutions for computing bisimulations of such large graphs\extended{~on commodity hardware.
One work using external memory for bisimulation is by Luo \etal}~\cite{DBLP:conf/cikm/LuoFHWB13}.
Our work is based on Apache Spark, so it is directly extensible to\extended{~be executed in both} a distributed computation environment of nodes with less main memory and/or using external memory\extended{~such as solid-state disks}.
}

\section{Conclusion and Future Work}
\label{sec:conclusion}

\response{We focus on the performance (runtime and memory use) of our generic, parallel \BRS algorithm for computing different bisimulation variants, and how this performance compares to specific algorithms for those bisimulations, due to Schätzle \etal and Kaushik \etal
Our experiments comparing $k=1, \dots, 10$ bisimulations on large synthetic and real-world graphs show that our generic, hash-based \BRS algorithm outperforms the respective native bisimulation algorithms on all datasets for all $k\geq5$ and for smaller~$k$ in some cases.
The experimental results indicate that the parallel \BRS algorithm and native Schätzle \etal scale linearly with the number of bisimulation iterations and the input graph's size.}
Our experiments also show that our generic \BRS-Kaushik algorithm effectively parallelizes the original sequential algorithm of Kaushik \etal, showing runtime performance similar to \BRS-Schätzle.
\extended{Our comparison is fair since both the generic \BRS algorithm and the reimplementations of the native algorithms were conducted using the same software stack (Scala and Apache Spark API).
We used a server machine with exclusive access during experimentation.
For very large graphs, an external memory solution is desirable.} %
Overall, we recommend using our generic \BRS algorithm over implementations of specific algorithms of $k$-bisimulations.
Due to the support of a formal language for defining graph summaries~\cite{DBLP:journals/tcs/BlumeRS21}, the \BRS approach is flexible and easily adaptable, \eg to changes in the desired features and user requirements, without sacrificing performance.

\response{Future work includes incorporating a check (similar to Kaushik \etal) for whether full bisimulation has been reached.
This is nontrivial as the algorithm of Kaushik \etal is based on splitting partitions, whereas \BRS is based on describing the equivalence class in which each vertex lives.
It is easy to check that no classes have been split, but harder to check that every pair of vertices that had the same description at the previous iteration have the same description at the current iteration, since the description of every vertex changes at each iteration.}

\extended{In additional experiments on an even larger Laundromat14B dataset with about $1.46$ billion vertices and $13.52$ billion edges, none of the algorithms could compute $10$-bisimulation within the $2\,\mathrm{TB}$ of memory available. %
We focus on performance comparison, so working in a large main memory of a single off-the-shelf machine is reasonable.
The results encourage us to use an external memory solution such as~\cite{DBLP:conf/cikm/LuoFHWB13} in future work.
The next step is to deploy our implementation in a distributed cloud-based compute environment with several smaller compute nodes and limited main memory.
Due to the use of Spark, the \BRS algorithm is well prepared to run in a distributed setting.
}

\paragraph{Acknowledgments}~This paper is the result of Jannik's Bachelor's thesis supervised by David and Ansgar.
We thank Till Blume for discussions and his support in implementing the $k$-bisimulation algorithms in FLUID~\cite{DBLP:conf/cikm/BlumeRS20}.

\bibliographystyle{splncs04} 
\bibliography{references}

\newpage


\appendix

\section*{Appendix}

\section{Additional Dataset Statistics}
\label{sec:appendix:datasets-stats}

Table~\ref{tab:statistics-datasets-degree} lists degree statistics of the datasets, where 
$d(G)$ is the average degree of vertices in $G$, 
$d_G(v)$ the degree of a vertex $v \in V$, and 
$\Delta(G)$ the maximum degree in $G$.

\begin{table*}[!h]
\small
        \begin{tabular}{|l|r@{$\,\pm\,$}l|r|r@{$\,\pm\,$}l|r|r@{$\,\pm\,$}l|r|}
        \hline
         \multicolumn{1}{|c|}{Graph} &
         \multicolumn{2}{c|}{$\mu(d_G(v))$} &
         \multicolumn{1}{c|}{$\Delta(G)$} &
         \multicolumn{2}{c|}{$\mu(d^-_G(v))$} &
         \multicolumn{1}{c|}{$\Delta^-(G)$} &
         \multicolumn{2}{c|}{$\mu(d^+_G(v))$} &
         \multicolumn{1}{c|}{$\Delta^+(G)$} \\
        \hline
        L'mat100M &  $5.89$ &    $569$ &  $1,570,748$ & $2.95$ & $559$ &  $1,570,748$ & $2.95$ & $108$ & $545,688$ \\
        BSBM100M       & $9.76$ &  $1,144$ &  $2,273,014$ & $5.04$ & $1144$ &  $2,273,014$ & $5.04$ & $6$ & $76$ \\
        BTC150M        & $58.43$ &  $5,655$ &  $5,629,275$ & $29.22$ & $504$ &    $283,686$ & $29.22$ & $5,629$ & $5,628,254$ \\
        BSBM1B         & $10.58$ &  $3,862$ & $23,924,441$ & $5.46$ & $3862$ & $23,924,441$ & $5.46$ & $6$ & $85$ \\ 
        BTC2B          & $48.40$ & $17,119$ & $65,879,409$ & $24.20$ & $1556$ &  $3,856,778$ & $24.20$ & $17,038$ & $65,878,298$ \\
        \hline
        \end{tabular}
    \caption{Degree Statistics of the Datasets.}
    \label{tab:statistics-datasets-degree}
\end{table*}

Laundromat100M has the smallest average and maximum values for both degree and in-degree.
BTC150M has the highest average and maximum degrees and the highest maximum in-degree. 
BSBM100M has the largest maximum in-degree and smallest maximum out-degree of our datasets.

We note that standard deviations in Table~\ref{tab:statistics-datasets-degree} are surprisingly large.
This is due to some vertices having degrees orders of magnitude larger than the average, as shown by the maximum degrees.

We note that, in the real-world datasets, average total degree is average in-degree plus average out-degree, to the precision quoted.
This is expected, because every edge adds one to the total degree of each of its endpoints.
This is not the case for the two versions of the synthetic BSBM dataset.
This dataset contains relatively many self-loops and, by convention, a self-loop adds $1$ to a vertex's total degree, while also adding $1$ to its in- and out-degrees.

\section{Results of the \BRS Algorithm}
\label{sec:appendix:breakdown-brs-with-initialization}
As the \BRS algorithm takes an additional initialization step (see Algorithm~\ref{alg:brs-adjusted}, lines \ref{algo:initial-signal}--\ref{algo:initial-signal-end}), Table~\ref{tab:brs-init-iter-runtime} shows the breakdown of the running time of the algorithms in initialization and computing the actual $k$ bisimulation iterations.
It is separated into average run time for initialization and computation of all ten iterations.

\begin{table}[ht!]

\small
        \centering
        \begin{tabular}{|l|r|r|r|r|}
        \hline
         & \multicolumn{2}{c|}{\BRS-Schätzle \etal} & \multicolumn{2}{c|}{\BRS-Kaushik \etal}  \\
         \cline{2-5}
         & \multicolumn{1}{c|}{Init.} & \multicolumn{1}{c|}{Iterations} & \multicolumn{1}{c|}{Init.} & \multicolumn{1}{c|}{Iterations} \\
        \hline        
        Laundromat100M & 1.10 & 4.46 & 1.10 & 4.50 \\
        BSBM100M & 1.00 & 5.46 & 1.05 & 4.15 \\
        BTC150M & 1.47 & 2.61 & 1.44 & 3.10 \\
        BSBM1B & 10.20 & 44.24 & 10.43 & 53.57  \\
        BTC2B & 16.22 & 45.74 & 16.29 & 69.63 \\
        \hline
    \end{tabular}
    \caption{Initialization and bisimulation iteration (minutes) of average total run time for \BRS.}
     \label{tab:brs-init-iter-runtime}

\end{table}

On all smaller datasets, \ie Laundromat100M, BSBM100M, and BTC150M and the initialization step of \BRS-Schätzle takes about $1$~minute on average (Table~\ref{tab:brs-init-iter-runtime}).
Regarding \BRS-Kaushik, the initialization also takes around $1$~minute on all smaller datasets (Table~\ref{tab:brs-init-iter-runtime}).
Regarding the larger dataset BSBM1B, the \BRS-Schätzle algorithm takes $10.20$ minutes for initialization and $44.24$ minutes for the $10$ iterations.
On BTC2B the algorithm needs $16.22$ minutes to initialize the graph and $45.74$ minutes to compute the $10$ iterations.
The \BRS-Kaushik takes on BSBM1B $10.43$ minutes for initialization and $53.57$ minutes for the $10$ iterations.
On BTC2B, the  \BRS-Kaushik algorithm needs $16.29$ minutes for initialization and $69.93$ minutes for the $10$ iterations.

Native Kaushik also has an initialization step, where the  vertices are partitioned based on their label.
This partitioning is done in parallel and took less than a second, so we do not account for it separately.

\section{Discussion of the Results of the Kaushik Algorithm}
\label{sec:app:detailed-discussion-kaushik}
Experimental results for native Kaushik indicate that the algorithm does not scale linearly with the input graph's size.
Its worst-case complexity is $\mathcal{O}(k \cdot m)$~\cite{DBLP:conf/icde/KaushikSBG02}, where $m$ is the number of edges in the input graph.

A reason for the long running time of native Kaushik on Laundromat100M could be the size of the initial partition~$P_0$.
It depends on the number of different label sets present in the graph's vertices (Definition~\ref{def:vertex-labeled-bw-bisimulation}).
Laundromat100M contains $33,431$ different label sets, which is much higher than for BTC150M with $69$ and BSBM100M with $1,289$ (Table~\ref{tab:statistics-datasets}).
As a consequence, the algorithm has to perform stability checks and (potential) splits on blocks more often than on the other datasets.
This does not contradict the similar run times on BTC150M and BSBM100M ($69$ label sets vs.\@ $1,289$ label sets).
First, BSBM100M reaches full bisimulation w.r.t.\@ Definition~\ref{def:vertex-labeled-bw-bisimulation} in iteration seven and hence execution is stopped early.
Second, as can be seen in Figure~\ref{plt:bsbm100M-log}, the iteration time starts to decrease rapidly from iteration four to five on BSBM100M, going down from about $20$~minutes to about $1$~minute.
This indicates that the partition hardly changes any further and therefore only a small number of stability checks and splits are performed in iterations five to seven.

Native Kaushik operates on the blocks of the graph's current partition.
In each iteration~$i$, the algorithm produces a partition~$P_i$, which is stable w.r.t.\@ all the blocks in~$P_{i-1}$~\cite{DBLP:conf/icde/KaushikSBG02}.
Consequently, if a block is split into two new blocks, these must also be checked for stability in that iteration (see also discussion in Section~\ref{sec:implementation}).
Hence, the more blocks are split in an iteration, the more steps must be performed by the algorithm.
In addition, bisimulation relationships between the vertices change far more often in early iterations \cite{DBLP:conf/sigmod/SchatzleNLP13}.
This explains the specific shape of the run time curve (Figures~\ref{plt:laundromat100-log}, \ref{plt:btc150M-log} and~\ref{plt:bsbm100M-log}) for the native Kaushik \etal algorithm.
The exception is BTC150M (Figure~\ref{plt:btc150M-log}), where the iteration time first decreases from iteration one to two, indicating that the $1$- and $2$-bisimulations of this graph are very similar.
 
\extended{
\section{Mappings of the notations used by Kaushik \etal and Paige and Tarjan}
\label{sec:appendix-mappings}

In \autoref{tab:notions} the notation used by Kaushik \etal, and by Paige and Tarjan is mapped to a single scheme.

\begin{table*}[t]
\small
    \begin{tabular}{|l|p{8cm}|M{2cm}|M{2cm}|}
        \hline
        Our notation & Meaning & Paige and Tarjan & Kaushik \etal\\
        \hline
        \hline
        $V$ & Finite set on which the relational coarsest partition will be computed (interpreted as a set of vertices in this work)  & $U$ & $V_G$ \\
        \hline
        $E \subseteq V \times V$ & Binary relation over the finite set $V$ (interpreted as a set of directed edges between the vertices in this work) & $E$ & $E_G$\\
        \hline
        $P_i$ and $P_i^{\cp}$ & Partition of $V$ and its copy in iteration $i \in \{0, 1, 2, \ldots\}$, where $P_0$ is the initial Partition & $P$ and $Q$ &$\mathcal{Q}$ and $\mathcal{X}$\\
        \hline
        $S \subseteq V$ & Any subset of the finite set $V$ & $S$ & $A$ and $B$ \\
        \hline
        $N^{+}(S) = \{y\in V \mid \exists x \in S \text{ with } xEy \}$ & The set of elements, which are related to any element $x \in S$ (if one interprets $V$ as vertices and $E$ as edges, then this refers to the out-neighbors of the vertices in $S$) & $E(S)$ & $Succ(A)$\\
        \hline
        $N^{-}(S) = \{x\in V \mid \exists y \in S \text{ with } xEy \}$ & The set of elements, which have any related element $y \in S$  (if one interprets $V$ as vertices and $E$ as edges, then this refers to the in-neighbors of the vertices in $S$) & $E^{-1}(S)$ & Not used \\
        \hline
        $B_{ij}$ & Block $j$ of Partition $P_i$ & $B$ & $Q$ and $X$\\
        \hline
    \end{tabular}
    \caption{Notation used in this work for the approach of Kaushik \etal}
    \label{tab:notions}
\end{table*}
}

\section{Detailed Results of Experiments}
\label{sec:appendix-results}

We provide the detailed experimental results.
%
\label{sec:appendix-results-small}
The results are split into results for the smaller datasets (Laundromat100M, BTC150M, BSBM100M) shown in \Cref{tab:detailed-results-laundromat100m,tab:detailed-results-btc150m,tab:detailed-results-bsbm100m}.
%
\label{sec:appendix-results-large}
The results for the larger datasets (BTC2B, BSBM1B) are shown in  \Cref{tab:detailed-results-btc2b,tab:detailed-results-bsbm1b}.
The tables for \BRS{} include an additional iteration $0$, which corresponds to the initialization routine of the algorithm.

\begin{table*}[h]
    \centering
    \subfloat[\BRS{} algorithm executed with GSM Schätzle ]{%
        \small
        \begin{tabular}{|*{15}{c|}}
        \hline
            \textbf{Run} & \multicolumn{11}{c|}{\textbf{Iteration}} & \multicolumn{3}{c|}{\textbf{Aggregates}} \\
            \hline
            \multicolumn{1}{|c|}{} & 0 & 1 & 2 & 3 & 4 & 5 & 6 & 7 & 8 & 9 & 10 & \textbf{$\Sigma$} & \textbf{$\mu$} & \textbf{$\sigma$} \\
            \hline 
            1 & 1.1 & 0.4 & 0.5 & 0.5 & 0.4 & 0.4 & 0.6 & 0.6 & 0.5 & 0.5 & 0.5 & 6.0 & 0.49 & 0.07 \\ 
            2 & 1.1 & 0.3 & 0.4 & 0.3 & 0.3 & 0.3 & 0.6 & 0.4 & 0.3 & 0.3 & 0.3 & 4.6 & 0.35 & 0.09 \\ 
            3 & 1.1 & 0.4 & 0.5 & 0.4 & 0.7 & 0.6 & 0.5 & 0.4 & 0.4 & 0.4 & 0.4 & 5.8 & 0.47 & 0.10 \\ 
            4 & 1.1 & 0.4 & 0.5 & 0.4 & 0.4 & 0.7 & 0.5 & 0.5 & 0.4 & 0.4 & 0.5 & 5.8 & 0.47 & 0.09 \\ 
            5 & 1.1 & 0.4 & 0.4 & 0.4 & 0.4 & 0.8 & 0.4 & 0.4 & 0.4 & 0.4 & 0.5 & 5.6 & 0.45 & 0.12 \\ 
            \hline
            \textbf{$\mu$} & 1.10 & 0.38 & 0.46 & 0.40 & 0.44 & 0.56 & 0.52 & 0.46 & 0.40 & 0.40 & 0.44 & \textbf{5.56} & \multicolumn{2}{c|}{} \\ 
            \hline
            \textbf{$\sigma$} & 0.00 & 0.04 & 0.05 & 0.06 & 0.14 & 0.19 & 0.07 & 0.08 & 0.06 & 0.06 & 0.08 & \textbf{0.50} & \multicolumn{2}{c|}{} \\
        \hline
        \end{tabular}
    }%
    \\
    \subfloat[Schätzle algorithm]{%
    \small
        \begin{tabular}{|*{14}{c|}}
        \hline
            \textbf{Run} & \multicolumn{10}{c|}{\textbf{Iteration}} & \multicolumn{3}{c|}{\textbf{Aggregates}} \\
            \hline
            \multicolumn{1}{|c|}{} & 1 & 2 & 3 & 4 & 5 & 6 & 7 & 8 & 9 & 10 & \textbf{$\Sigma$} & \textbf{$\mu$} & \textbf{$\sigma$} \\
            \hline
            1 & 1.1 & 0.8 & 0.7 & 0.8 & 0.8 & 0.7 & 0.7 & 0.7 & 0.7 & 0.8 & 7.8 & 0.78 & 0.12 \\ 
            2 & 1.1 & 0.8 & 0.7 & 0.8 & 0.7 & 0.7 & 0.7 & 0.7 & 0.7 & 0.7 & 7.6 & 0.76 & 0.12 \\ 
            3 & 0.7 & 0.8 & 1.1 & 0.8 & 0.7 & 0.8 & 0.8 & 0.7 & 0.7 & 0.7 & 7.8 & 0.78 & 0.12 \\ 
            4 & 0.7 & 0.8 & 1.1 & 0.8 & 0.7 & 0.8 & 0.7 & 0.7 & 0.7 & 0.7 & 7.7 & 0.77 & 0.12 \\ 
            5 & 0.7 & 0.7 & 1.1 & 0.8 & 0.7 & 0.8 & 0.8 & 0.7 & 0.7 & 0.7 & 7.7 & 0.77 & 0.12 \\ 
            \hline
            \textbf{$\mu$} & 0.86 & 0.78 & 0.94 & 0.80 & 0.72 & 0.76 & 0.74 & 0.70 & 0.70 & 0.72 & \textbf{7.72} & \multicolumn{2}{c|}{} \\
            \hline
            \textbf{$\sigma$} & 0.00 & 0.00 & 0.00 & 0.08 & 0.04 & 0.07 & 0.00 & 0.00 & 0.05 & 0.04 & \textbf{0.07} & \multicolumn{2}{c|}{} \\
        \hline
        \end{tabular}
    }%
    \\
    \subfloat[\BRS{} algorithm executed with GSM Kaushik]{%
    \small
        \begin{tabular}{|*{15}{c|}}
        \hline
            \textbf{Run} & \multicolumn{11}{c|}{\textbf{Iteration}} & \multicolumn{3}{c|}{\textbf{Aggregates}} \\
             \hline
             \multicolumn{1}{|c|}{} & 0 & 1 & 2 & 3 & 4 & 5 & 6 & 7 & 8 & 9 & 10 & \textbf{$\Sigma$} & \textbf{$\mu$} & \textbf{$\sigma$} \\
            \hline 
            1 & 1.1 & 0.3 & 0.4 & 0.4 & 0.4 & 0.4 & 0.4 & 0.6 & 0.7 & 0.5 & 0.5 & 5.7 & 0.46 & 0.11 \\ 
            2 & 1.1 & 0.3 & 0.4 & 0.3 & 0.3 & 0.4 & 0.5 & 0.7 & 0.4 & 0.4 & 0.5 & 5.3 & 0.42 & 0.12 \\ 
            3 & 1.1 & 0.3 & 0.4 & 0.4 & 0.5 & 0.7 & 0.5 & 0.5 & 0.4 & 0.4 & 0.5 & 5.7 & 0.46 & 0.10 \\ 
            4 & 1.1 & 0.3 & 0.4 & 0.4 & 0.4 & 0.4 & 0.7 & 0.6 & 0.5 & 0.4 & 0.5 & 5.7 & 0.46 & 0.11 \\ 
            5 & 1.1 & 0.3 & 0.4 & 0.4 & 0.4 & 0.7 & 0.4 & 0.5 & 0.4 & 0.4 & 0.6 & 5.6 & 0.45 & 0.11 \\ 
            \hline
            \textbf{$\mu$} & 1.10 & 0.30 & 0.40 & 0.38 & 0.40 & 0.52 & 0.50 & 0.58 & 0.48 & 0.42 & 0.52 & \textbf{5.60} & \multicolumn{2}{c|}{} \\
            \hline
            \textbf{$\sigma$} & 0.00 & 0.00 & 0.00 & 0.04 & 0.06 & 0.15 & 0.11 & 0.07 & 0.12 & 0.04 & 0.04 & \textbf{0.15} & \multicolumn{2}{c|}{} \\
        \hline
        \end{tabular}
    }%
    \\
    \subfloat[Kaushik algorithm]{%
        \small
        \begin{tabular}{|*{14}{c|}}
        \hline
            \textbf{Run} & \multicolumn{10}{c|}{\textbf{Iteration}} & \multicolumn{3}{c|}{\textbf{Aggregates}} \\
             \hline
             \multicolumn{1}{|c|}{} & 1 & 2 & 3 & 4 & 5 & 6 & 7 & 8 & 9 & 10 & \textbf{$\Sigma$} & \textbf{$\mu$} & \textbf{$\sigma$} \\
            \hline
            1 & 11.4 & 21.2 & 35.1 & 60.0 & 85.5 & 106.7 & 98.8 & 81.8 & 59.3 & 44.1 & 603.9 & 60.39 & 30.9 \\ 
            2 & 13.6 & 22.7 & 35.8 & 66.8 & 76.8 & 91.9 & 85.6 & 69.6 & 52.1 & 35.0 & 549.9 & 54.99 & 25.81 \\ 
            3 & 10.9 & 19.3 & 31.9 & 53.7 & 81.3 & 100.1 & 98.7 & 79.9 & 62.5 & 42.5 & 580.8 & 58.08 & 30.21 \\ 
            4 & 11.2 & 21.1 & 33.3 & 58.2 & 78.6 & 100.5 & 98.1 & 85.1 & 61.6 & 41.2 & 588.9 & 58.89 & 30.04 \\ 
            5 & 11.8 & 21.3 & 34.1 & 55.1 & 81.8 & 107.7 & 97.0 & 91.9 & 69.0 & 40.1 & 609.8 & 60.98 & 31.81 \\
            \hline
            \textbf{$\mu$} & 11.78 & 21.12 & 34.04 & 58.76 & 80.80 & 101.38 & 95.64 & 81.66 & 60.90 & 40.58 & \textbf{586.66} & \multicolumn{2}{c|}{} \\
            \hline
            \textbf{$\sigma$} & 0.96 & 1.08 & 1.37 & 4.59 & 2.97 & 5.67 & 5.06 & 7.28 & 5.45 & 3.09 & \textbf{21.09} & \multicolumn{2}{c|}{} \\
        \hline
        \end{tabular}
    }%
    \\
    \caption{Detailed Results (minutes) on Laundromat100M for $10$-bisimulation. }
    \label{tab:detailed-results-laundromat100m}
\end{table*}

\begin{table*}[h]
    \centering
    \subfloat[\BRS{} algorithm executed with GSM Schätzle ]{%
        \small
        \begin{tabular}{|*{15}{c|}}
        \hline
            \textbf{Run} & \multicolumn{11}{c|}{\textbf{Iteration}} & \multicolumn{3}{c|}{\textbf{Aggregates}} \\
            \hline
            \multicolumn{1}{|c|}{} & 0 & 1 & 2 & 3 & 4 & 5 & 6 & 7 & 8 & 9 & 10 & \textbf{$\Sigma$} & \textbf{$\mu$} & \textbf{$\sigma$} \\
            \hline 
            1 & 1.4 & 0.1 & 0.1 & 0.1 & 0.1 & 0.1 & 0.1 & 0.1 & 0.1 & 0.1 & 0.1 & 2.4 & 0.10 & 0.00 \\ 
            2 & 1.4 & 0.4 & 0.4 & 0.4 & 0.8 & 0.5 & 0.4 & 0.4 & 0.4 & 0.5 & 0.4 & 6.0 & 0.46 & 0.12 \\ 
            3 & 1.7 & 0.2 & 0.1 & 0.1 & 0.1 & 0.1 & 0.1 & 0.1 & 0.1 & 0.1 & 0.6 & 3.3 & 0.16 & 0.15 \\ 
            4 & 1.4 & 0.2 & 0.1 & 0.1 & 0.1 & 0.1 & 0.1 & 0.1 & 0.1 & 0.1 & 0.4 & 2.8 & 0.14 & 0.09 \\ 
            5 & 1.7 & 0.4 & 0.4 & 0.4 & 0.6 & 0.4 & 0.4 & 0.4 & 0.4 & 0.4 & 0.4 & 5.9 & 0.42 & 0.06 \\  
            \hline
            \textbf{$\mu$} & 1.47 & 0.26 & 0.22 & 0.22 & 0.34 & 0.24 & 0.22 & 0.22 & 0.22 & 0.24 & 0.38 & \textbf{4.08} & \multicolumn{2}{c|}{} \\
            \hline
            \textbf{$\sigma$} & 0.14 & 0.12 & 0.15 & 0.15 & 0.3 & 0.17 & 0.15 & 0.15 & 0.15 & 0.17 & 0.16 & \textbf{1.55} & \multicolumn{2}{c|}{} \\
        \hline
        \end{tabular}
    }%
    \\
    \subfloat[Schätzle algorithm]{%
        \small
        \begin{tabular}{|*{14}{c|}}
        \hline
            \textbf{Run} & \multicolumn{10}{c|}{\textbf{Iteration}} & \multicolumn{3}{c|}{\textbf{Aggregates}} \\
             \hline
             \multicolumn{1}{|c|}{} & 1 & 2 & 3 & 4 & 5 & 6 & 7 & 8 & 9 & 10 & \textbf{$\Sigma$} & \textbf{$\mu$} & \textbf{$\sigma$} \\
            \hline
            1 & 0.6 & 0.6 & 0.7 & 0.8 & 0.7 & 0.6 & 0.6 & 0.6 & 0.6 & 0.6 & 6.4 & 0.64 & 0.07 \\ 
            2 & 0.5 & 0.6 & 0.7 & 0.8 & 0.6 & 0.6 & 0.6 & 0.6 & 0.6 & 0.6 & 6.2 & 0.62 & 0.07 \\ 
            3 & 0.6 & 0.6 & 0.8 & 0.8 & 0.6 & 0.6 & 0.6 & 0.6 & 0.6 & 0.6 & 6.4 & 0.64 & 0.08 \\ 
            4 & 0.5 & 0.6 & 0.8 & 0.8 & 0.6 & 0.6 & 0.6 & 0.6 & 0.6 & 0.6 & 6.3 & 0.63 & 0.09 \\ 
            5 & 0.5 & 0.5 & 0.6 & 0.7 & 0.6 & 0.5 & 0.5 & 0.5 & 0.5 & 0.5 & 5.4 & 0.54 & 0.07 \\ 
            \hline
            \textbf{$\mu$} & 0.54 & 0.58 & 0.72 & 0.78 & 0.62 & 0.58 & 0.58 & 0.58 & 0.58 & 0.58 & \textbf{6.14} & \multicolumn{2}{c|}{} \\
            \hline
            \textbf{$\sigma$} & 0.05 & 0.04 & 0.07 & 0.04 & 0.04 & 0.04 & 0.04 & 0.04 & 0.04 & 0.04 & \textbf{0.38} & \multicolumn{2}{c|}{} \\
        \hline
        \end{tabular}
    }%
    \\
    \subfloat[\BRS{} algorithm executed with GSM Kaushik ]{%
        \small
        \begin{tabular}{|*{15}{c|}}
        \hline
            \textbf{Run} & \multicolumn{11}{c|}{\textbf{Iteration}} & \multicolumn{3}{c|}{\textbf{Aggregates}} \\
            \hline
            \multicolumn{1}{|c|}{} & 0 & 1 & 2 & 3 & 4 & 5 & 6 & 7 & 8 & 9 & 10 & \textbf{$\Sigma$} & \textbf{$\mu$} & \textbf{$\sigma$} \\
            \hline 
            1 & 1.4 & 0.4 & 0.4 & 0.3 & 0.4 & 0.6 & 0.4 & 0.4 & 0.4 & 0.4 & 0.4 & 5.5 & 0.41 & 0.07 \\ 
            2 & 1.4 & 0.1 & 0.1 & 0.1 & 0.1 & 0.1 & 0.1 & 0.1 & 0.1 & 0.1 & 0.7 & 3.0 & 0.16 & 0.18 \\ 
            3 & 1.4 & 0.1 & 0.1 & 0.1 & 0.1 & 0.1 & 0.1 & 0.1 & 0.1 & 0.1 & 0.5 & 2.8 & 0.14 & 0.12 \\ 
            4 & 1.4 & 0.4 & 0.5 & 0.5 & 0.4 & 0.4 & 0.4 & 0.4 & 0.4 & 0.4 & 0.4 & 5.6 & 0.42 & 0.04 \\ 
            5 & 1.4 & 0.4 & 0.6 & 0.5 & 0.4 & 0.4 & 0.4 & 0.4 & 0.4 & 0.4 & 0.5 & 5.8 & 0.44 & 0.07 \\ 
            \hline
            \textbf{$\mu$} & 1.44 & 0.28 & 0.34 & 0.30 & 0.28 & 0.32 & 0.28 & 0.28 & 0.28 & 0.28 & 0.50 & \textbf{4.54} & \multicolumn{2}{c|}{} \\
            \hline
            \textbf{$\sigma$} & 0.14 & 0.15 & 0.21 & 0.18 & 0.15 & 0.19 & 0.15 & 0.15 & 0.15 & 0.15 & 0.11 & \textbf{1.34} & \multicolumn{2}{c|}{} \\
        \hline
        \end{tabular}
    }%
    \\
    \subfloat[Kaushik algorithm]{%
        \small
        \begin{tabular}{|*{14}{c|}}
        \hline
            \textbf{Run} & \multicolumn{10}{c|}{\textbf{Iteration}} & \multicolumn{3}{c|}{\textbf{Aggregates}} \\
            \hline
            \multicolumn{1}{|c|}{} & 1 & 2 & 3 & 4 & 5 & 6 & 7 & 8 & 9 & 10 & \textbf{$\Sigma$} & \textbf{$\mu$} & \textbf{$\sigma$} \\
            \hline
            1 & 3.9 & 0.9 & 3.7 & 12.2 & 17.0 & 17.8 & 12.7 & 6.6 & 2.8 & 1.0 & 78.6 & 7.86 & 6.17 \\ 
            2 & 3.8 & 0.9 & 3.7 & 11.2 & 16.8 & 17.0 & 12.9 & 6.0 & 2.9 & 1.5 & 76.7 & 7.67 & 5.92 \\ 
            3 & 4.1 & 0.9 & 3.9 & 11.7 & 17.6 & 17.6 & 12.6 & 6.1 & 3.1 & 1.1 & 78.7 & 7.87 & 6.14 \\ 
            4 & 3.9 & 0.9 & 4.0 & 14.1 & 20.5 & 18.5 & 12.2 & 6.0 & 2.5 & 1.2 & 83.8 & 8.38 & 6.95 \\ 
            5 & 3.6 & 0.9 & 3.5 & 10.8 & 15.5 & 16.0 & 12.0 & 6.3 & 2.7 & 1.0 & 72.3 & 7.23 & 5.55 \\ 
            \hline
            \textbf{$\mu$} & 3.86 & 0.90 & 3.76 & 12.00 & 17.48 & 17.38 & 12.48 & 6.20 & 2.80 & 1.16 & \textbf{78.02} & \multicolumn{2}{c|}{} \\
            \hline
            \textbf{$\sigma$} & 0.16 & 0.00 & 0.17 & 1.15 & 1.66 & 0.84 & 0.33 & 0.23 & 0.20 & 0.19 & \textbf{3.71} & \multicolumn{2}{c|}{} \\
        \hline
        \end{tabular}
    }%
    \\
    \caption{Detailed Results (minutes) on BTC150M for $10$-bisimulation. }
    \label{tab:detailed-results-btc150m}
\end{table*}

\begin{table*}[h]
    \centering
    \subfloat[\BRS{} algorithm executed with GSM Schätzle ]{%
        \small
        \begin{tabular}{|*{15}{c|}}
        \hline
            \textbf{Run} & \multicolumn{11}{c|}{\textbf{Iteration}} & \multicolumn{3}{c|}{\textbf{Aggregates}} \\
            \hline
            \multicolumn{1}{|c|}{} & 0 & 1 & 2 & 3 & 4 & 5 & 6 & 7 & 8 & 9 & 10 & \textbf{$\Sigma$} & \textbf{$\mu$} & \textbf{$\sigma$} \\
            \hline 
            1 & 1.0 & 0.5 & 0.5 & 0.5 & 0.6 & 0.8 & 0.5 & 0.5 & 0.5 & 0.5 & 0.4 & 6.3 & 0.53 & 0.10 \\ 
            2 & 1.0 & 0.5 & 0.6 & 0.5 & 0.5 & 0.9 & 0.5 & 0.5 & 0.5 & 0.5 & 0.5 & 6.5 & 0.55 & 0.12 \\ 
            3 & 1.0 & 0.5 & 0.6 & 0.5 & 0.5 & 0.9 & 0.5 & 0.5 & 0.5 & 0.5 & 0.5 & 6.5 & 0.55 & 0.12 \\ 
            4 & 1.0 & 0.5 & 0.5 & 0.5 & 0.6 & 0.9 & 0.5 & 0.5 & 0.5 & 0.5 & 0.5 & 6.5 & 0.55 & 0.12 \\ 
            5 & 1.0 & 0.5 & 0.5 & 0.5 & 0.9 & 0.5 & 0.5 & 0.5 & 0.5 & 0.6 & 0.5 & 6.5 & 0.55 & 0.12 \\ 
            \hline
            \textbf{$\mu$} & 1.00 & 0.50 & 0.54 & 0.50 & 0.62 & 0.80 & 0.50 & 0.50 & 0.50 & 0.52 & 0.48 & \textbf{6.46} & \multicolumn{2}{c|}{} \\
            \hline
            \textbf{$\sigma$} & 0.00 & 0.00 & 0.05 & 0.00 & 0.15 & 0.15 & 0.00 & 0.00 & 0.00 & 0.04 & 0.04 & \textbf{0.08} & \multicolumn{2}{c|}{} \\
        \hline
        \end{tabular}
    }%
    \\
    \subfloat[Schätzle algorithm]{%
        \small
        \begin{tabular}{|*{14}{c|}}
        \hline
            \textbf{Run} & \multicolumn{10}{c|}{\textbf{Iteration}} & \multicolumn{3}{c|}{\textbf{Aggregates}} \\
            \hline
            \multicolumn{1}{|c|}{} & 1 & 2 & 3 & 4 & 5 & 6 & 7 & 8 & 9 & 10 & \textbf{$\Sigma$} & \textbf{$\mu$} & \textbf{$\sigma$} \\
            \hline
            1 & 0.9 & 1.3 & 0.9 & 0.9 & 0.9 & 0.9 & 0.9 & 0.9 & 0.9 & 0.9 & 9.4 & 0.94 & 0.12 \\ 
            2 & 0.9 & 1.0 & 1.2 & 0.9 & 1.0 & 0.9 & 0.9 & 0.9 & 0.9 & 0.9 & 9.5 & 0.95 & 0.09 \\ 
            3 & 0.9 & 0.9 & 1.2 & 0.9 & 0.9 & 0.9 & 0.9 & 0.9 & 0.9 & 0.9 & 9.3 & 0.93 & 0.09 \\ 
            4 & 0.9 & 1.0 & 1.2 & 0.9 & 0.9 & 0.9 & 0.9 & 0.9 & 0.9 & 0.9 & 9.4 & 0.94 & 0.09 \\ 
            5 & 0.9 & 1.0 & 1.2 & 0.9 & 0.9 & 0.9 & 0.9 & 0.9 & 0.9 & 0.9 & 9.4 & 0.94 & 0.09 \\   
            \hline
            \textbf{$\mu$} & 0.90 & 1.04 & 1.14 & 0.90 & 0.92 & 0.90 & 0.90 & 0.90 & 0.90 & 0.90 & \textbf{9.40} & \multicolumn{2}{c|}{} \\
            \hline
            \textbf{$\sigma$} & 0.00 & 0.14 & 0.12 & 0.00 & 0.04 & 0.00 & 0.00 & 0.00 & 0.00 & 0.00 & \textbf{0.06} & \multicolumn{2}{c|}{} \\
        \hline
        \end{tabular}
    }%
    \\
    \subfloat[\BRS{} algorithm executed with GSM Kaushik ]{%
        \small
        \begin{tabular}{|*{15}{c|}}
        \hline
            \textbf{Run} & \multicolumn{11}{c|}{\textbf{Iteration}} & \multicolumn{3}{c|}{\textbf{Aggregates}} \\
            \hline
            \multicolumn{1}{|c|}{} & 0 & 1 & 2 & 3 & 4 & 5 & 6 & 7 & 8 & 9 & 10 & \textbf{$\Sigma$} & \textbf{$\mu$} & \textbf{$\sigma$} \\
            \hline 
            1 & 1.0 & 0.4 & 0.5 & 0.4 & 0.4 & 0.6 & 0.7 & 0.4 & 0.4 & 0.4 & 0.4 & 5.6 & 0.46 & 0.10 \\ 
            2 & 1.0 & 0.3 & 0.4 & 0.3 & 0.3 & 0.3 & 0.3 & 0.5 & 0.5 & 0.3 & 0.3 & 4.5 & 0.35 & 0.08 \\ 
            3 & 1.1 & 0.4 & 0.4 & 0.4 & 0.5 & 0.7 & 0.5 & 0.4 & 0.4 & 0.4 & 0.5 & 5.7 & 0.46 & 0.09 \\ 
            4 & 1.0 & 0.4 & 0.5 & 0.4 & 0.4 & 0.6 & 0.6 & 0.4 & 0.4 & 0.4 & 0.4 & 5.5 & 0.45 & 0.08 \\ 
            5 & 1.1 & 0.3 & 0.4 & 0.3 & 0.3 & 0.3 & 0.3 & 0.5 & 0.5 & 0.3 & 0.4 & 4.7 & 0.36 & 0.08 \\ 
            \hline
            \textbf{$\mu$} & 1.05 & 0.36 & 0.44 & 0.36 & 0.38 & 0.50 & 0.48 & 0.44 & 0.44 & 0.36 & 0.40 & \textbf{5.20} & \multicolumn{2}{c|}{}\\
            \hline
            \textbf{$\sigma$} & 0.05 & 0.05 & 0.05 & 0.05 & 0.07 & 0.17 & 0.16 & 0.05 & 0.05 & 0.05 & 0.06 & \textbf{0.50} & \multicolumn{2}{c|}{}\\
        \hline
        \end{tabular}
    }%
    \\
    \subfloat[Kaushik algorithm]{%
        \small
        \begin{tabular}{|*{14}{c|}}
        \hline
            \textbf{Run} & \multicolumn{10}{c|}{\textbf{Iteration}} & \multicolumn{3}{c|}{\textbf{Aggregates}} \\
            \hline
            \multicolumn{1}{|c|}{} & 1 & 2 & 3 & 4 & 5 & 6 & 7 & 8 & 9 & 10 & \textbf{$\Sigma$} & \textbf{$\mu$} & \textbf{$\sigma$} \\
            \hline
            1 & 8.0 & 16.5 & 33.9 & 19.0 & 0.8 & 0.6 & 0.6 & 0.0 & 0.0 & 0.0 & 79.4 & 7.94 & 11.03 \\ 
            2 & 8.8 & 17.2 & 29.3 & 22.1 & 0.9 & 0.6 & 0.6 & 0.0 & 0.0 & 0.0 & 79.5 & 7.95 & 10.44 \\ 
            3 & 8.2 & 16.2 & 30.7 & 22.6 & 0.8 & 0.7 & 0.7 & 0.0 & 0.0 & 0.0 & 79.9 & 7.99 & 10.71 \\ 
            4 & 8.0 & 19.0 & 26.4 & 21.5 & 0.8 & 0.6 & 0.6 & 0.0 & 0.0 & 0.0 & 76.9 & 7.69 & 9.97 \\ 
            5 & 9.2 & 15.9 & 25.8 & 20.6 & 0.8 & 0.6 & 0.6 & 0.0 & 0.0 & 0.0 & 73.5 & 7.35 & 9.43 \\ 
            \hline
            \textbf{$\mu$} & 8.44 & 16.96 & 29.22 & 21.16 & 0.82 & 0.62 & 0.62 & 0.00 & 0.00 & 0.00 & \textbf{77.84} & \multicolumn{2}{c|}{} \\
            \hline
            \textbf{$\sigma$} & 0.48 & 1.11 & 2.96 & 1.27 & 0.04 & 0.04 & 0.04 & 0.00 & 0.00 & 0.00 & \textbf{2.41} & \multicolumn{2}{c|}{} \\
        \hline
        \end{tabular}
    }%
    \\
    \caption{Detailed Results (minutes) on BSBM100M for $10$-bisimulation. }
    \label{tab:detailed-results-bsbm100m}
\end{table*}

\begin{table*}[h]
    \centering
    \subfloat[\BRS{} algorithm executed with GSM Schätzle ]{%
    \small
        \begin{tabular}{|*{15}{c|}}
        \hline
            \textbf{Run} & \multicolumn{11}{c|}{\textbf{Iteration}} & \multicolumn{3}{c|}{\textbf{Aggregates}} \\
            \hline
            \multicolumn{1}{|c|}{} & 0 & 1 & 2 & 3 & 4 & 5 & 6 & 7 & 8 & 9 & 10 & \textbf{$\Sigma$} & \textbf{$\mu$} & \textbf{$\sigma$} \\
            \hline 
            1 & 16.0 & 4.7 & 5.2 & 4.9 & 5.0 & 5.6 & 5.1 & 5.1 & 5.1 & 5.1 & 5.9 & 67.7 & 5.17 & 0.33 \\
            2 & 16.0 & 4.5 & 5.1 & 4.9 & 5.0 & 5.9 & 5.0 & 5.0 & 5.8 & 5.1 & 5.3 & 67.6 & 5.16 & 0.40 \\ 
            3 & 16.4 & 1.6 & 1.8 & 1.7 & 1.9 & 2.2 & 1.9 & 1.9 & 1.9 & 1.9 & 6.0 & 39.2 & 2.28 & 1.25 \\ 
            4 & 16.0 & 4.5 & 5.1 & 4.9 & 5.2 & 5.5 & 5.1 & 5.0 & 5.8 & 5.2 & 5.4 & 67.7 & 5.17 & 0.33 \\ 
            5 & 16.4 & 4.6 & 4.9 & 4.8 & 6.2 & 5.7 & 5.0 & 5.0 & 4.9 & 5.0 & 5.1 & 67.6 & 5.12 & 0.45 \\ 
            \hline
            \textbf{$\mu$} & 16.22 & 3.98 & 4.42 & 4.24 & 4.66 & 4.98 & 4.42 & 4.40 & 4.70 & 4.46 & 5.54 & \textbf{61.96} & \multicolumn{2}{c|}{}\\
            \hline
            \textbf{$\sigma$} & 0.16 & 1.19 & 1.31 & 1.27 & 1.45 & 1.4 & 1.26 & 1.25 & 1.45 & 1.28 & 0.35 & \textbf{11.38} & \multicolumn{2}{c|}{}\\
        \hline
        \end{tabular}
    }%
    \\
    \subfloat[Schätzle algorithm]{%
        \small
        \begin{tabular}{|*{14}{c|}}
        \hline
            \textbf{Run} & \multicolumn{10}{c|}{\textbf{Iteration}} & \multicolumn{3}{c|}{\textbf{Aggregates}} \\
            \hline
            \multicolumn{1}{|c|}{} & 1 & 2 & 3 & 4 & 5 & 6 & 7 & 8 & 9 & 10 & \textbf{$\Sigma$} & \textbf{$\mu$} & \textbf{$\sigma$} \\
            \hline
            1 & 8.3 & 8.7 & 8.7 & 9.2 & 8.7 & 8.7 & 8.7 & 8.7 & 8.7 & 8.9 & 87.3 & 8.73 & 0.21 \\ 
            2 & 8.1 & 8.3 & 8.3 & 8.8 & 8.3 & 8.3 & 8.3 & 8.3 & 8.3 & 8.3 & 83.3 & 8.33 & 0.17 \\ 
            3 & 7.9 & 8.2 & 8.2 & 8.2 & 8.1 & 8.2 & 8.1 & 8.6 & 8.1 & 8.2 & 81.8 & 8.18 & 0.17 \\ 
            4 & 7.9 & 8.3 & 8.3 & 8.4 & 8.4 & 8.7 & 8.8 & 8.4 & 8.4 & 8.5 & 84.1 & 8.41 & 0.23 \\ 
            5 & 7.9 & 8.2 & 8.1 & 8.2 & 8.3 & 8.3 & 8.3 & 8.3 & 8.3 & 8.3 & 82.2 & 8.22 & 0.12 \\  
            \hline
            \textbf{$\mu$} & 8.02 & 8.34 & 8.32 & 8.56 & 8.36 & 8.44 & 8.44 & 8.46 & 8.36 & 8.44 & \textbf{83.74} & \multicolumn{2}{c|}{} \\
            \hline
            \textbf{$\sigma$} & 0.16 & 0.19 & 0.20 & 0.39 & 0.20 & 0.22 & 0.27 & 0.16 & 0.20 & 0.25 & \textbf{1.96} & \multicolumn{2}{c|}{} \\
        \hline
        \end{tabular}
    }%
    \\
    \subfloat[\BRS{} algorithm executed with GSM Kaushik ]{%
    \small
        \begin{tabular}{|*{15}{c|}}
        \hline
            \textbf{Run} & \multicolumn{11}{c|}{\textbf{Iteration}} & \multicolumn{3}{c|}{\textbf{Aggregates}} \\
            \hline
            \multicolumn{1}{|c|}{} & 0 & 1 & 2 & 3 & 4 & 5 & 6 & 7 & 8 & 9 & 10 & \textbf{$\Sigma$} & \textbf{$\mu$} & \textbf{$\sigma$} \\
            \hline 
            1 & 16.1 & 1.9 & 2.5 & 3.8 & 7.0 & 7.0 & 7.5 & 7.2 & 7.4 & 7.2 & 7.3 & 74.9 & 5.88 & 2.11 \\ 
            2 & 16.1 & 2.1 & 2.7 & 4.2 & 7.0 & 8.3 & 7.5 & 8.0 & 7.6 & 8.0 & 11.8 & 83.3 & 6.72 & 2.77 \\ 
            3 & 16.6 & 2.4 & 2.8 & 4.3 & 7.9 & 7.8 & 8.0 & 7.8 & 7.8 & 7.8 & 10.9 & 84.1 & 6.75 & 2.55 \\ 
            4 & 16.1 & 1.9 & 2.6 & 3.9 & 7.1 & 7.3 & 7.7 & 7.2 & 7.3 & 7.1 & 7.1 & 75.3 & 5.92 & 2.10 \\ 
            5 & 16.6 & 5.2 & 5.9 & 7.4 & 11.2 & 11.1 & 11.2 & 10.7 & 10.6 & 10.8 & 11.3 & 112.0 & 9.54 & 2.28 \\ 
            \hline
            \textbf{$\mu$} & 16.29 & 2.70 & 3.30 & 4.72 & 8.04 & 8.30 & 8.38 & 8.18 & 8.14 & 8.18 & 9.68 & \textbf{85.92} & \multicolumn{2}{c|}{}\\
            \hline
            \textbf{$\sigma$} & 0.20 & 1.26 & 1.30 & 1.35 & 1.62 & 1.47 & 1.42 & 1.30 & 1.24 & 1.35 & 2.05 & \textbf{13.60} & \multicolumn{2}{c|}{}\\
        \hline
        \end{tabular}
    }%
    \\
    \caption{Detailed Results (minutes) on BTC2B for $10$-bisimulation. }
    \label{tab:detailed-results-btc2b}
\end{table*}

\begin{table*}[h]
    \centering
    \subfloat[\BRS{} algorithm executed with GSM Schätzle ]{%
        \small
        \begin{tabular}{|*{15}{c|}}
        \hline
            \textbf{Run} & \multicolumn{11}{c|}{\textbf{Iteration}} & \multicolumn{3}{c|}{\textbf{Aggregates}} \\
            \hline
            \multicolumn{1}{|c|}{} & 0 & 1 & 2 & 3 & 4 & 5 & 6 & 7 & 8 & 9 & 10 & \textbf{$\Sigma$} & \textbf{$\mu$} & \textbf{$\sigma$} \\
            \hline 
            1 & 10.1 & 3.9 & 4.7 & 4.3 & 4.3 & 4.5 & 5.3 & 4.7 & 5.1 & 4.7 & 5.7 & 57.3 & 4.72 & 0.50 \\ 
            2 & 10.1 & 2.9 & 3.7 & 3.3 & 3.4 & 3.3 & 4.1 & 4.1 & 4.1 & 3.8 & 3.1 & 45.9 & 3.58 & 0.42 \\ 
            3 & 10.0 & 3.9 & 4.8 & 4.3 & 4.4 & 4.5 & 5.2 & 4.7 & 4.9 & 4.5 & 4.2 & 55.4 & 4.54 & 0.36 \\ 
            4 & 10.1 & 3.9 & 4.8 & 4.4 & 4.4 & 4.5 & 5.2 & 4.9 & 5.2 & 4.5 & 5.8 & 57.7 & 4.76 & 0.51 \\ 
            5 & 10.0 & 3.9 & 4.8 & 4.4 & 4.4 & 4.6 & 5.4 & 4.8 & 5.0 & 4.4 & 4.2 & 55.9 & 4.59 & 0.41 \\ 
            \hline
            \textbf{$\mu$} & 10.20 & 3.70 & 4.56 & 4.14 & 4.18 & 4.28 & 5.04 & 4.64 & 4.86 & 4.38 & 4.60 & \textbf{54.44} & \multicolumn{2}{c|}{}\\
            \hline
            \textbf{$\sigma$} & 0.62 & 0.40 & 0.43 & 0.42 & 0.39 & 0.49 & 0.48 & 0.28 & 0.39 & 0.31 & 1.02 & \textbf{4.35} & \multicolumn{2}{c|}{}\\
        \hline
        \end{tabular}
    }%
    \\
    \subfloat[Schätzle algorithm]{%
    \small
        \begin{tabular}{|*{14}{c|}}
        \hline
            \textbf{Run} & \multicolumn{10}{c|}{\textbf{Iteration}} & \multicolumn{3}{c|}{\textbf{Aggregates}} \\
            \hline
            \multicolumn{1}{|c|}{} & 1 & 2 & 3 & 4 & 5 & 6 & 7 & 8 & 9 & 10 & \textbf{$\Sigma$} & \textbf{$\mu$} & \textbf{$\sigma$} \\
            \hline
            1 & 9.0 & 8.1 & 8.1 & 8.2 & 8.3 & 8.2 & 8.2 & 8.5 & 8.2 & 8.2 & 83.0 & 8.30 & 0.26 \\ 
            2 & 10.0 & 8.3 & 8.2 & 8.3 & 8.4 & 8.5 & 8.4 & 8.4 & 8.3 & 8.3 & 85.1 & 8.51 & 0.50 \\ 
            3 & 8.8 & 8.9 & 8.9 & 9.0 & 9.0 & 12.5 & 9.3 & 9.0 & 9.0 & 9.0 & 93.4 & 9.34 & 1.06 \\ 
            4 & 8.1 & 8.2 & 8.4 & 8.2 & 8.3 & 8.6 & 8.7 & 8.2 & 8.2 & 8.2 & 83.1 & 8.31 & 0.19 \\ 
            5 & 8.5 & 8.4 & 8.3 & 8.4 & 8.5 & 9.2 & 8.5 & 8.5 & 8.5 & 8.5 & 85.3 & 8.53 & 0.23 \\  
            \hline
            \textbf{$\mu$} & 8.88 & 8.38 & 8.38 & 8.42 & 8.50 & 9.40 & 8.62 & 8.52 & 8.44 & 8.44 & \textbf{85.98} & \multicolumn{2}{c|}{} \\
            \hline
            \textbf{$\sigma$} & 0.64 & 0.28 & 0.28 & 0.30 & 0.26 & 1.58 & 0.38 & 0.26 & 0.30 & 0.30 & \textbf{3.83} & \multicolumn{2}{c|}{} \\
        \hline
        \end{tabular}
    }%
    \\
    \subfloat[\BRS{} algorithm executed with GSM Kaushik ]{%
        \small
        \begin{tabular}{|*{15}{c|}}
        \hline
            \textbf{Run} & \multicolumn{11}{c|}{\textbf{Iteration}} & \multicolumn{3}{c|}{\textbf{Aggregates}} \\
            \hline
            \multicolumn{1}{|c|}{} & 0 & 1 & 2 & 3 & 4 & 5 & 6 & 7 & 8 & 9 & 10 & \textbf{$\Sigma$} & \textbf{$\mu$} & \textbf{$\sigma$} \\
            \hline 
            1 & 10.7 & 4.0 & 5.6 & 5.8 & 5.9 & 6.0 & 5.8 & 5.8 & 5.8 & 6.3 & 5.0 & 66.7 & 5.60 & 0.62 \\ 
            2 & 10.7 & 3.2 & 5.2 & 5.5 & 5.4 & 5.5 & 5.5 & 5.5 & 5.4 & 6.3 & 5.6 & 63.8 & 5.31 & 0.75 \\ 
            3 & 10.3 & 3.1 & 5.0 & 5.2 & 5.2 & 5.1 & 5.1 & 5.2 & 5.2 & 6.0 & 4.0 & 59.4 & 4.91 & 0.76 \\ 
            4 & 10.7 & 3.2 & 5.1 & 5.2 & 5.5 & 5.4 & 5.3 & 5.3 & 5.3 & 5.7 & 5.1 & 61.8 & 5.11 & 0.66 \\ 
            5 & 10.3 & 4.1 & 5.8 & 6.1 & 6.1 & 6.1 & 6.1 & 6.1 & 6.1 & 6.6 & 4.9 & 68.3 & 5.80 & 0.70 \\ 
            \hline
            \textbf{$\mu$} & 10.43 & 3.52 & 5.34 & 5.56 & 5.62 & 5.62 & 5.56 & 5.58 & 5.56 & 6.18 & 4.92 & \textbf{64.00} & \multicolumn{2}{c|}{}\\
            \hline
            \textbf{$\sigma$} & 0.27 & 0.44 & 0.31 & 0.35 & 0.33 & 0.38 & 0.36 & 0.33 & 0.34 & 0.31 & 0.52 & \textbf{3.22} & \multicolumn{2}{c|}{}\\
        \hline
        \end{tabular}
    }%
    \\
    \caption{Detailed Results (minutes) on BSBM1B for $10$-bisimulation. }
    \label{tab:detailed-results-bsbm1b}
\end{table*}

\end{document}